\documentclass[iicol]{sn-jnl}% Default with double column layout

\usepackage[T1]{fontenc}
\usepackage{palatino}
\usepackage{microtype}
\usepackage{algorithm} 
\usepackage{algpseudocode} 
\usepackage{color}

\newtheorem{theorem}{Theorem}
\newtheorem{proposition}[theorem]{Proposition}
\newtheorem{corollary}[theorem]{Corollary}
\newtheorem{lemma}[theorem]{Lemma}
\newtheorem{remark}{Remark}
\newtheorem{definition}{Definition}

\usepackage{graphicx,epstopdf,epsfig}
\usepackage{array}
\usepackage{multicol}
\usepackage{multirow}
\usepackage[caption=false,font=normalsize,labelfont=sf,textfont=sf]{subfig}
\usepackage{textcomp}
\usepackage{stfloats}
\usepackage{url}
\usepackage{verbatim}
\usepackage{relsize}

\def\i{\boldsymbol{i}}
\def\j{\boldsymbol{j}}

\def\x{\boldsymbol{x}}

\def\v{\boldsymbol{v}}
\def\y{\boldsymbol{y}}

\def\z{\boldsymbol{z}}

\def\b{\boldsymbol{b}}

\def\u{\boldsymbol{u}}
\def\v{\boldsymbol{v}}

\def\btau{\boldsymbol{\tau}}
\def\A{\mathbf{A}}
\def\W{\mathbf{W}}
\def\J{\mathcal{J}}
\def\B{\mathbf{B}}

\def\I{\mathrm{I}}
\def\X{\mathbf{X}}
\def\Y{\mathbf{Y}}

\def\D{\mathrm{D}}
\def\N{\mathrm{N}}
\def\F{\mathbf{F}}
\def\G{\mathrm{G}}
\def\T{\mathrm{T}}

\def\R{\mathbb{R}}
\def\S{\mathbf{S}}
\def\U{\mathrm{U}}

\def\V{\mathrm{V}}

\def\Pr{\mathrm{P}}
\def\Pcal{\mathcal{P}}
\def\bxi{\boldsymbol{\xi}}
\def\boldeta{\boldsymbol{\eta}}
\def\prox{\mathrm{Prox}}

\def\FBSop{{\mathrm{FBS}}}

\def\DRSop{{\mathrm{DRS}}}

\newcommand*{\tri}{\Lambda}
\newcommand{\norm}[1]{\left\lVert#1\right\rVert}

\DeclareMathOperator*{\argmin}{\arg\!\min}

\renewcommand{\Re}{\mathbb{R}}

\begin{document}

\title[Article Title]{Averaged Deep Denoisers for Image Regularization}
\author*[1]{\fnm{Pravin} \sur{Nair}}\email{pravinn@iisc.ac.in}
\author[1]{\fnm{Kunal N.} \sur{Chaudhury}}\email{kunal@iisc.ac.in}

\affil*[1]{\orgdiv{Department of Electrical Engineering}, \orgname{Indian Institute of Science}, \orgaddress{\city{Bangalore},  \country{India}}}

\abstract{Plug-and-Play (PnP) and Regularization-by-Denoising (RED) are recent paradigms for image reconstruction that leverage the power of modern denoisers for image regularization. In particular, they have been shown to deliver state-of-the-art reconstructions with CNN denoisers. Since the regularization is performed in an ad-hoc manner, understanding the convergence of PnP and RED has been an active research area. It was shown in recent works that iterate convergence can be guaranteed if the denoiser is averaged or nonexpansive. However, integrating nonexpansivity with gradient-based learning is challenging, the core issue being that testing nonexpansivity is intractable. Using numerical examples, we show that existing CNN denoisers tend to violate the nonexpansive property, which can cause PnP or RED to diverge. In fact, algorithms for training nonexpansive denoisers either cannot guarantee nonexpansivity or are computationally intensive. In this work, we construct contractive and averaged image denoisers by unfolding splitting-based optimization algorithms applied to wavelet denoising and demonstrate that their regularization capacity for PnP and RED can be matched with CNN denoisers. To our knowledge, this is the first work to propose a simple framework for training contractive denoisers using network unfolding.}

 
\maketitle

\section{Introduction}

The problem of recovering an unknown image from incomplete, corrupted measurements comes up in applications such as deblurring, superresolution, inpainting, demosaicking, despeckling, MRI, and tomography \cite{ribes2008linear, engl1996regularization,dong2011image}. The standard optimization framework for such problems involves a loss function $f: \Re^n \to \Re$ derived from the measurement process and a regularizer $g: \Re^n \to \Re \cup \{\infty\}$ incorporating prior knowledge about the ground truth. The reconstruction is performed by solving the  optimization problem
\begin{equation}
\label{mainopt}
\underset{\x \in \Re^n}{\min} \  f(\x) + g(\x).  
\end{equation}
The loss $f$ is typically differentiable, while the regularizer $g$ can be discontinuous \cite{engl1996regularization}. For linear inverse problems such as deblurring, superresolution, inpainting, and compressed sensing \cite{dong2011image,jagatap2019algorithmic}, the loss is convex and quadratic, namely, $f(\x)={\lVert \y - \A\x \rVert}^2$, where $\A \in \Re^{m \times n}$ is the measurement matrix and $\y$ is the observed image. On the other hand, for applications such as single photon imaging, Poisson denoising, and despeckling  \cite{chan2017plug,rond2016poisson,bioucas2010multiplicative}, $f$ is non-quadratic. 

For imaging applications, total-variation and other sparsity-promoting functionals \cite{rudin1992nonlinear,candes2008enhancing,dong2011sparsity} are the most commonly used regularizers. For such regularizers, $g$ is typically convex but nondifferentiable \cite{engl1996regularization,zhang2017image}. If the loss $f$ is convex, we can solve \eqref{mainopt} using iterative algorithms such as Forward-Backward Splitting (FBS), Douglas-Rachford Splitting (DRS), and Alternating Direction Method of Multipliers (ADMM) \cite{passty1979ergodic,bauschke2017convex,ryu2016primer}. Model-based optimization is interpretable, computationally tractable, and can be applied to various problems. However, handcrafted regularizers are generally less powerful than modern learning-based models, and the optimization process can be time-consuming \cite{zhang2017image}.

\subsection{PnP and RED}

Classical algorithms such as FBS and DRS perform repeated inversion of the measurement model, followed by a regularization step.
%The present work is motivated by plug-and-play algorithms, where the regularization is implicitly performed within a proximal algorithm. 
More precisely, the regularization is performed using the proximal operator. For a convex regularizer $h: \Re^n \to \Re$, the proximal operator    
is defined as
\begin{equation}
\label{prox}
\prox_{h}(\x) = \argmin_{\z \in \Re^n}\  \left\{\frac{1}{2} \norm{\z - \x}^2 +  h(\z)\right\},
\end{equation}
where $\| \cdot \|$ is the Euclidean norm. The proximal operator can be seen as a denoiser --- it takes a noisy image $\x$ and returns an image close to $\x$ for which the penalty $h$ is small \cite{beck2009fast}. This operator is at the core of FBS, DRS, and ADMM.
In particular, the FBS update is given by
\begin{align}
\label{fbs}
{\z}_{k+1} &= {\x}_{k} -  \gamma \nabla \! f ({\x}_{k}), \nonumber \\
{\x}_{k+1} &= \prox_{\gamma g} (\z_{k+1}). 
\end{align}
where $\gamma > 0$  and $\x_0 \in \Re^n$ is an arbitrary initialization. 

On the other hand, starting with $\x_0  \in \Re^n$ and $\gamma > 0$, the DRS update is given by (e.g., see \cite{ryu2016primer})
\begin{align}
\label{drs}
{\z}_{k+1} &= \prox_{\gamma f}(\x_k),\nonumber \\
{\x}_{k+1} &= \x_k+\prox_{\gamma g} (2\z_{k+1}-\x_k)-\z_{k+1}. 
\end{align}

%ADMM is simply DRS applied to the Fenchel conjugate of a constrained formulation of \eqref{mainopt}. We will not detail the ADMM updates since we do not use it in this work.

%We observe from \eqref{fbs} and \eqref{drs} that FBS is cheaper than the DRS if computing $\nabla \! f$ is less expensive than computing its proximal operator. In fact, the advantage with DRS is that the loss $f$ need not be differentiable; however, the proximal operator of $f$ should be efficiently computable, either in closed form or iteratively \cite{parikh2014proximal}. 

%\begin{algorithm}
%	\caption{PNP-DRS} 
%	\begin{algorithmic}[1]
%		\State \textbf{input}:  $\x_0$, loss $f$,  denoiser $\D$,  and $\gamma >0$.	
%		\For {$k=0,1,\ldots$}
%			\State $\z_{k+1} = \D (\x_k)$.
%			\State $\x_{k+1} =  \x_k+ \D (2\z_{k+1}-\x_k)-\z_{k+1}$.
%		\EndFor
%	\end{algorithmic} 
%	\label{pnp-drs} 
%\end{algorithm}

In  Plug-and-Play (PnP) regularization \cite{sreehari2016plug}, instead of having to handcraft the regularizer $g$ in \eqref{mainopt} and apply it via its proximal operator, we directly ``plug''  a powerful denoising operator $\D: \R^n \to \R^n$, such as  BM3D \cite{dabov2007image} or DnCNN \cite{zhang2017beyond}, into \eqref{fbs} and \eqref{drs}. Thus, in the PnP variant of FBS (referred to as PnP-FBS), the update is performed by replacing the proximal operator in \eqref{fbs} by an image denoiser  (Algorithm \ref{pnp-fbs}).
%\begin{equation}
%\label{pnpista}
%{\z}_{k+1} = {\x}_{k} -  \rho \nabla \! f ({\x}_{k}), \ {\x}_{k+1} = \D (\z_{k+1}).
%\end{equation}
%Similarly, in the PnP variant of DRS (PnP-DRS),  $ \prox_{\gamma g} $ in \eqref{pnp-fbs} is replace by denoiser $\D$ (Algorithm \ref{pnp-drs}).
%\begin{align}
%\label{pnpadmm}
%& \v_{k+1} = \D (\x_k - \z_k), \\ 
%& \x_{k+1} = \prox_{\rho f} (\v_{k+1} + \z_k), \nonumber \\ 
%& \z_{k+1} = \z_k + (\v_{k+1} - \x_{k+1}). \nonumber
%\end{align}

\begin{algorithm}
	\caption{PNP-FBS} 
	\begin{algorithmic}[1]
		\State \textbf{input}: $\x_0$, loss $f$, denoiser $\D$, and $\gamma >0$.	
		\For {$k=0,1,\ldots$}
			\State $\z_{k+1} = {\x}_{k} -  \gamma \nabla \! f ({\x}_{k})$.
			\State $\x_{k+1}= \D (\z_{k+1})$ .
		\EndFor
	\end{algorithmic} 
\label{pnp-fbs}	
\end{algorithm}

Remarkably, PnP algorithms yield state-of-the-art results across a wide range of applications, including superresolution, MRI, fusion, and tomography \cite{sreehari2016plug,Ryu2019_PnP_trained_conv,Sun2019_PnP_SGD,zhang2017learning,Tirer2019_iter_denoising,zhang2021plug}.
The best performing PnP algorithms \cite{zhang2017learning,Tirer2019_iter_denoising,zhang2021plug} use trained denoisers such as DnCNN \cite{zhang2017beyond}, IRCNN \cite{zhang2017learning}, and U-Net \cite{hurault2022gradient,hurault2022proximal}. This also opens up the exciting possibility of combining model-based reconstruction with deep learning \cite{zhang2017learning}. 

The idea of using denoisers for regularization has also been explored in Regularization-by-Denoising (RED). For example, the authors in \cite{romano2017little} proposed to work with the regularizer
\begin{equation}
\label{redregularizer} 
g(\x) = \frac{1}{2} \x^\top\big(\x - \D(\x)\big),
\end{equation}
%and the corresponding reconstruction problem
%\begin{equation}
%\label{mainoptimred}
%\underset{\x}{\mathrm{argmin}} \  \left\{f(\x) + \frac{1}{2}  \x^\top\big(\x - \D(\x)\big) \right\}, 
%\end{equation}
where $\D$ is some powerful denoiser as in PnP. This is solved iteratively using gradient-based methods, where the gradient of  \eqref{redregularizer} is approximated using $\x - \D(\x)$ in \cite{romano2017little}. It was later shown in \cite{reehorst2018regularization} that this is exact if $\D$ is locally homogenous and its Jacobian is symmetric.

Several variants of the original RED algorithm \cite{romano2017little} have been proposed \cite{reehorst2018regularization,Sun2019_PnP_SGD,sun2019block,sun2021scalable,cohen2021regularization}. One such variant called RED-PG \cite{reehorst2018regularization} is described in Algorithm \ref{red-pg}. RED algorithms have been shown to yield state-of-the-art results for phase retrieval, tomography, superresolution, deblurring, and compressed sensing \cite{metzler2018prdeep,wu2020simba,wu2019online,mataev2019deepred,sun2019block,hu2022monotonically}. 

\begin{algorithm}
	\caption{RED-PG} 
	\begin{algorithmic}[1]
		\State \textbf{input}:  $\z_0$, loss $f$,  denoiser $\D$,  and $\gamma, L >0$.	
		\For {$k=0,1,\ldots$}
			\State $\x_{k+1} =\prox_{(\gamma /L)f } (\z_k)$.
			\State $\z_{k+1} = \frac{1}{L} \big(\D(\x_{k+1}) + (L-1)\x_{k+1}\big)$.
		\EndFor
	\end{algorithmic} 
	\label{red-pg} 
\end{algorithm}

There is yet another paradigm for solving inverse problems called deep unfolding \cite{gregor2010learning,sun2016deep,zhang2018ista}, where an iterative algorithm is unfolded into a network and trained in an end-to-end fashion. 
%The main idea behind these so-called unfolded networks is to interpret a truncated iterative algorithm as an end-to-end trainable deep network, where the intrinsic parameters of the algorithm are learned from the data and where each iteration is interpreted as a layer. 
Deep unfolded networks can produce impressive results, but unlike PnP or RED (where just the denoiser needs to be trained), we have to train an unfolded network end-to-end. The present work is motivated by deep unfolding, but we propose to use them for a different purpose, namely, to construct nonexpansive denoisers by unfolding classical wavelet denoising. We then apply the trained denoiser within Algorithms \ref{pnp-fbs} and \ref{red-pg} for image reconstruction. 

We wish to point to a related work \cite{repetti2022dual}, where the authors construct an unfolded denoiser by applying FBS to a sparsity-regularized image denoising problem (with box constraints on the reconstruction). However, there are important distinctions between our approach and theirs. First, splitting is applied to the Fenchel dual of the denoising problem in \cite{repetti2022dual}; the idea is to transform the nonsmooth primal (denoising) problem into a smooth, unconstrained dual problem. Second, since the sparsifying linear transform is optimized in \cite{repetti2022dual}, there is no structural guarantee on the trained layers. As a result, it is difficult to come up with convergence guarantees; indeed, \cite[Proposition~2]{repetti2022dual} assumes that the same linear transform is used across all layers, and this would generally not hold if we optimize the transforms. We present comparisons with \cite{repetti2022dual} in Section~\ref{Experiments}.

\subsection{Convergence Results}

The success of PnP and RED has sparked interest in the theoretical understanding of these algorithms. For example, it has been shown that for a class of linear denoisers, PnP is guaranteed to converge, even at a linear rate \cite{ACK2023-contractivity}, and the limit point is the solution of a convex optimization problem \cite{sreehari2016plug,nair2021fixed,gavaskar2021plug}. This is done by expressing the linear denoiser as the proximal operator of a convex function. Unfortunately, this does not hold for powerful nonlinear denoisers such as BM3D, DnCNN, or IRCNN. This is because, unlike the proximal operator that is nonexpansive, such denoisers have been shown to violate the nonexpansive property \cite{reehorst2018regularization}.

Though developing a variational characterization for deep denoisers is challenging, it is nonetheless possible to establish iterate convergence for PnP and RED algorithms. For example, the convergence of Algorithm \ref{pnp-fbs} was established for generative denoisers in \cite{jagatap2019algorithmic,liu2021recovery,raj2019gan}. For strongly convex loss functions, PnP-FBS convergence can be guaranteed using a CNN denoiser whose residual is nonexpansive \cite{Ryu2019_PnP_trained_conv}. Motivated by RED, convergence was recently established for PnP modeled on FBS, ADMM, half-quadratic splitting, and gradient descent \cite{hurault2022gradient,cohen2021has,hurault2022proximal}. On the other hand, the convergence of RED algorithms can be guaranteed using a nonexpansive denoiser \cite{romano2017little,reehorst2018regularization,sun2019block,cohen2021regularization,liu2021recovery}. 
%More specifically, along with nonexpansivity, the denoiser in \cite{romano2017little} must be locally homogenous and its Jacobian must be symmetric.
For instance, the block-coordinate RED algorithm in \cite{sun2019block} exhibits convergence with a nonexpansive denoiser. In \cite{liu2021recovery}, the authors established convergence of RED for the compressed sensing application using a nonexpansive denoiser. 

The focus of this work is on strong convergence of the PnP and RED algorithms,  where the iterates converge in norm to a fixed point of a well-defined operator \cite{ryu2016primer}. It is important to note that there exist other forms of iterate convergence. For example, for the denoisers in \cite{hurault2022gradient,hurault2022proximal}, the limit points of the PnP-FBS iterates are stationary points of a nonconvex objective function. In contrast, the iterates converge weakly  (see \cite[Section $2.5$]{bauschke2017convex} for a definition) to a zero of a combination of a gradient operator and a firmly nonexpansive operator in \cite{pesquet2021learning}. On the other hand, the iterates of Plug-and-Play Stochastic Gradient Descent (PnP-SGD) in \cite{laumont2023maximum} converge to a stationary point of a MAP problem, where the prior is derived from the denoiser.    
%In \cite{cohen2021regularization}, RED is formulated as a convex optimization problem using a nonexpansive denoiser.  
%As against these, a new variant of RED is proposed in \cite{hu2022monotonically}, where the denoiser is not required to be nonexpansive for algorithmic convergence. 
%We note that the connection between PnP and RED algorithms has been explored in \cite{liu2021recovery,cohen2021regularization}; this allows us to transfer the convergence results of one algorithm to the other.
The summary of our understanding of the iterate convergence of PnP and RED is as follows:
\begin{enumerate}
\item Convergence of PnP is guaranteed for averaged denoisers \cite{Sun2019_PnP_SGD,sun2021scalable,nair2021fixed}.

\item For nonexpansive denoisers, convergence is guaranteed for variants of RED   \cite{reehorst2018regularization,sun2019block,cohen2021regularization}, PnP-FBS applied to compressive sensing \cite{liu2021recovery}, and a variant of PnP-FBS \cite{reehorst2018regularization}.
\end{enumerate}

For the above reasons, there has been significant research on training averaged and nonexpansive CNN denoisers. The challenge is that computing the  Lipschitz constant of a neural network is NP-hard \cite{virmaux2018lipschitz}. By Lipschitz constant, we mean the smallest $c$ such that $\lVert \T(\u) -\T(\v) \rVert \leqslant c \lVert \u-\v \rVert$ for all inputs $\u$ and $\v$, where we represent the CNN as an operator $\T$. In particular, while the Lipschitz constant of a convolution layer is just the largest singular value of the matrix representation of the layer (which can be computed efficiently from the FFT of its filters \cite{sedghi2018singular}) and the Lipschitz constant of common nonlinear activations (ReLU and sigmoid) is $1$, computing the Lipschitz constant for their cascaded composition is however nontrivial. Nonetheless, we can compute an upper bound for the Lipschitz constant using a direct approximation for the spectral norm of the network  \cite{pesquet2021learning} or the product of the spectral norms of the convolution layers \cite{Ryu2019_PnP_trained_conv}. In particular, if we force the spectral norm of the convolution layers to be within $1$ at each epoch of the training process, then the final network is automatically nonexpansive. Several methods based on this idea have been proposed \cite{sedghi2018singular,hertrich2021convolutional,terris2020building}. In particular, we will later compare with the nonexpansive CNN denoiser in \cite{Ryu2019_PnP_trained_conv}, where the largest singular value of a convolutional layer (estimated using power iterations) is used to force the Lipschitz constant of the layer to be within $1$ after each gradient update.

\subsection{Motivation} 
\label{motivation}

This work was motivated by the observation that existing CNN denoisers tend to violate the nonexpansive property (see Fig.~\ref{nonexpansiveplots}). In fact, existing methods for training nonexpansive denoisers either cannot guarantee the final network to be nonexpansive or are computationally expensive. For example, power iterations are used for approximate computation of the largest singular value in \cite{Ryu2019_PnP_trained_conv,pesquet2021learning}; hence, the network is not guaranteed to be nonexpansive at each epoch. The methods in  \cite{sedghi2018singular,hertrich2021convolutional,terris2020building} for constraining the convolutional layers are computationally expensive; e.g., the time taken in \cite{sedghi2018singular} is about nine-fold that required to train the network without constraints. This is because \cite{sedghi2018singular} requires the SVD computation of a large matrix derived from the Fourier transform of the convolutional filters. Similarly, \cite{terris2020building} requires the Fourier transform of the filters in every convolutional layer. On the other hand, the training in \cite{hertrich2021convolutional} requires us to solve an optimization problem that imposes orthogonality constraints on the convolutional layers.

We also highlight a technical issue that has been sidelined in existing works \cite{Ryu2019_PnP_trained_conv,sedghi2018singular,hertrich2021convolutional,terris2020building,pesquet2021learning}, namely, that the CNN filters are trained on images of fixed size but deployed on images of arbitrary size. Since the filters must be zero-padded to match the image size, their Lispchitz can change with the size. As a concrete example, consider the $3 \times 3$  Sobel filter $[1, 0, -1; 2, 0, -2; 1,  0, -1]$. The Lipschitz constant of this filter for an image of size $25 \times 25$ is $7.9842$, whereas the Lipschitz constant for a $256 \times 256$ image is $8$. Thus, a guarantee that the Lipschitz bound is independent of the image size is not available for existing CNN  denoisers. To substantiate our point, we present numerical evidence (see Fig.~\ref{nonexpansiveplots} in Section \ref{Experiments}) showing that BM3D \cite{dabov2007image}, DnCNN  \cite{zhang2017beyond}, and N-CNN \cite{Ryu2019_PnP_trained_conv} violate the nonexpansive property. Such counterexamples have also been reported in  \cite{reehorst2018regularization}.  Our observations are summarized below:
\begin{enumerate}

\item The Lipschitz constant of a CNN (trained with fixed-size filters) depends on the size of the input image and is difficult to predict.
 
\item Existing CNN denoisers violate the nonexpansive property, which can result in the divergence of PnP and RED algorithms (see Tables \ref{counterexample1} and \ref{counterexample2} in Section \ref{Experiments}).

\item Developing efficient algorithms for training nonexpansive CNNs is challenging. Existing algorithms for training nonexpansive CNNs either cannot guarantee nonexpansivity or are computationally expensive.

\end{enumerate}

\subsection{Contributions}

Following the above observations, we moved away from CNNs and considered deep unfolded networks instead. In fact, from our experience, we noticed a tradeoff between nonexpansivity and the denoising performance of CNNs --- a significant drop in denoising performance was observed if the denoiser was forced to be nonexpansive. Indeed, it was shown in \cite{hurault2022proximal,hertrich2021convolutional} that imposing hard Lipschitz constraints on the CNN can adversely affect its denoising performance. 

In the rest of the paper, we demonstrate that it is possible to construct deep unfolded networks that come with the following desirable properties:
\begin{enumerate}
\item, Unlike \cite{Ryu2019_PnP_trained_conv,sedghi2018singular}, they are averaged (or contractive) by construction and hence automatically nonexpansive. 
\item The training is not expensive compared to  \cite{sedghi2018singular,terris2020building,pesquet2021learning}. In particular, we need to only perform cheap projections onto intervals  (of the learnable parameters) after every gradient step.
\item When plugged into Algorithms \ref{pnp-fbs} and \ref{red-pg}, the reconstruction is comparable with the state-of-the-art. 
\end{enumerate}

Our construction is motivated by the averaged (contractivity) property of fixed-point operators arising in FBS and DRS; these are classically used to establish iterate convergence \cite{bauschke2017convex}. In fact, the building blocks of our denoiser are obtained by unfolding FBS and DRS applied to wavelet denoising. 
These blocks are provably averaged (resp.~contractive), and the proposed denoiser, being a composition of these building blocks, is automatically averaged (resp.~contractive). While our denoiser is trained on small patches, we give a formula for patch aggregation that guarantees the final end-to-end image denoiser to be averaged (or contractive). The denoising capacity of our denoiser is short of BM3D and DnCNN; however, their regularization capacity for PnP and RED is comparable with BM3D and DnCNN. The source code is available in \url{https://github.com/pravin1390/Averageddeepdenoiser}.

\section{Notations and Definitions}
\label{notationsec}

For integers $p \leqslant q$, we will use the notations $[p,q]=\{p,p+1,\ldots,q\}$ and $[p,q]^2=[p,q] \times [p,q]$. We also use  $[p,q]_s$ to denote the integers $\{ps,(p+1)s,\ldots, qs\}$, which are multiples of $s$.

We will work in $\Re^n$ and the norm $\lVert \, \cdot \, \rVert$ will be the Euclidean norm. We say that a sequence $\{\x_k\}$ in $\Re^n$ converges to $\x^* \in \Re^n$ if $\lVert \x_k - \x^* \rVert \to 0$ as $k \to \infty$. Moreover, the convergence is linear if there exists $C >0$ and $\mu \in (0,1)$ such that $\lVert \x_k - \x^* \rVert \leqslant C \mu^k$. A sequence is said to be convergent if it converges to some $\x^* \in \Re^n$; it is said to converge to a set $S \subset \Re^n$ if it converges to some $\x^* \in S$.  A sequence  $\{\x_k\}$ generated by an iterative algorithm is said to converge globally if it converges for any arbitrary initialization $\x_0 \in \Re^n$.

We recall the definitions and properties of certain nonlinear operators that come up in the fixed-point convergence of PnP and RED algorithms. These will play an important role in the construction of the proposed denoiser.

An operator $\T: \Re^n \to \Re^n$ is said to be Lipschitz  if there exists $c >0$ such that
\begin{equation}
\label{defLip}
\forall \, \x,\y \in \Re^n: \quad \norm{\T (\x) - \T(\y)} \leqslant c \norm{\x - \y}.
\end{equation}
The smallest $c$ satisfying \eqref{defLip} is called the Lipschitz constant of $\T$. An operator $\T$ is nonexpansive if $c=1$ and contractive if $c \in [0,1)$. $\T$ is said to be $\theta$-averaged for some $\theta \in [0,1)$, if we can write $\T = (1-\theta) \, \I + \theta \, \N$, where $\N$ is a nonexpansive operator and $\I$ is the identity operator on $\Re^n$. We call $\T$ firmly nonexpansive if $\T$ is $(1/2)$-averaged.

\vspace{-2mm}

\begin{remark}
It is evident from the above definitions that an averaged operator is nonexpansive. The converse is not true; e.g., the operator $-\I$ is nonexpansive but not averaged. A contractive operator is $(c+1)/2$-averaged, where $c$ is the contraction factor. However, not every averaged operator is contractive. Thus, contractive operators form a strict subset of averaged operators, which form a strict subset of nonexpansive operators \cite{bauschke2017convex}. 
\end{remark}
\vspace{-3mm}

Two instances of averaged operators relevant to this work are the proximal and the gradient-step operators. Recall that for a convex function $h: \Re^n \to \Re$, its proximal operator $\prox_h: \R^n \to \R^n$ is defined in \eqref{prox}. 
%\vspace{-3mm}
%\begin{lemma}
%\label{lem:halfaveraged_prox}
%Let $h : \Re^n \to \Re$ be convex and $\alpha >0$. Then $\prox_{\alpha h}$ in \eqref{prox} is firmly nonexpansive.
%\end{lemma}
%\vspace{-3mm}
%
For a differentiable function $f: \Re^n \to \Re$, we define the gradient-step operator to be 
\begin{equation}
\label{gradstepop}
\G_{\gamma,f}: \Re^n \to \Re^n, \quad \G_{\gamma,f} = \I - \gamma \nabla \! f,
\end{equation}
where $\gamma > 0$ is the step size.  A differentiable function $f$ is said to be $\beta$-smooth if the Lipschitz constant of the gradient operator $\nabla \! f: \Re^n \to \Re^n$ is at most $\beta$. 

\vspace{-3mm}
\begin{remark}
The proximal and the gradient-step operators in \eqref{prox}  and  \eqref{gradstepop} are known to be firmly nonexpansive when $f$ is convex \cite{bauschke2017convex}. However, these operators need not be contractive unless additional conditions are met, e.g., strong convexity of $f$. However, for linear inverse problems, the loss $f$ is typically not strongly convex, and it is easy to show that the proximal and the gradient-step operators are not contractive in this case. This will be an important consideration when we work with proximal and gradient-step operators.
\end{remark}
\vspace{-3mm}

\section{Convergence of PnP and RED}
\label{sec:convg}

The convergence analysis of operator-splitting methods is primarily based on the classical fixed-point theory of monotone operators \cite{bauschke2017convex,ryu2016primer,parikh2014proximal,eckstein1992douglas}. PnP and RED have their roots in convex optimization and operator-splitting algorithms in particular. It is thus not surprising that existing convergence analyses of PnP and RED rely heavily on the fixed-point theory of firmly nonexpansive and averaged operators. \cite{nair2021fixed,sun2021scalable,Ryu2019_PnP_trained_conv,cohen2021regularization,reehorst2018regularization}. In particular, this line of analysis can be used to establish convergence of Algorithms \ref{pnp-fbs} and \ref{red-pg}. This is done by expressing the updates in these algorithms as a mapping $\x \mapsto \T(\x)$, where $\T$ is contractive or averaged depending on the loss $f$ and the denoiser $\D$. In particular, consider the operators
\begin{equation}
\label{PnPop}
\T_{\mathrm{PnP}} = \D \circ \G_{\gamma,f}, 
\end{equation}
and
\begin{equation}
\label{REDop}
\T_{\mathrm{RED}} =\prox_{(\gamma /L)f }  \circ  \frac{1}{L} \big(\D + (L-1) \, \I \, \big).
\end{equation}
In terms \eqref{PnPop} and \eqref{REDop}, we can express the updates in Algorithms \ref{pnp-fbs} and \ref{red-pg} as $\x_{k+1}=\T(\x_k)$, where the operator $\T$ is either \eqref{PnPop} or \eqref{REDop}. Namely, we can view PnP-FBS and RED-PG as fixed-point algorithms. 

The convergence analysis of fixed-point iterations is a classical topic and of research interest in the area of image regularization \cite{FPTbook}. In particular, since operators \eqref{PnPop} and \eqref{REDop} are continuous, a limit point of the iterates (if one exists) must necessarily be a fixed point of $\T$; we call this fixed-point convergence. We will denote the fixed points of $\T$ as $\mathrm{Fix} \, \T$, i.e., $\mathrm{Fix} \, \T = \{\x \in \R^n: \ \x=\T(\x) \}$.

We can establish various convergence results of the fixed-point iterations of $\T$. % depending on the nature of $\T$ and the notion of convergence. 
A classical result is the following. 

\vspace{-3mm}
\begin{theorem}
\label{thm:Banach}
Let $\T$ be a contractive operator on $\R^n$. Then the sequence $\{\x_k\}_{k \geqslant 0}$ generated as  $\x_{k+1} = \T(\x_k)$ converges globally and linearly to $\mathrm{Fix} \, \T$.
\end{theorem}
\vspace{-3mm}

The iteration $\x_{k+1} = \T(\x_k)$ is called the Picard iteration when $\T$ is contractive, and Theorem~\ref{thm:Banach} is the contraction-mapping theorem \cite{FPTbook}.

%By ``linear'' convergence, we mean that the distance of $\x_k$ from  $\mathrm{Fix} \, \T$ goes to zero exponentially in $k$.

\vspace{-3mm}
\begin{remark}
$\mathrm{Fix} \, \T$ is nonempty if $\T$ is contractive. In fact, $\mathrm{Fix} \, \T$ is a singleton in this case, i.e., the fixed-point iterations always converge to the same fixed point regardless of the initialization $\x_0$. In general, however, verifying that $\mathrm{Fix} \, \T \neq \emptyset$ for some operator $\T$ is a nontrivial task; this is assumed in most practical applications of fixed-point analysis, including the analysis of operator-splitting algorithms \cite{bauschke2017convex,ryu2016primer}. 
\end{remark}
\vspace{-3mm}

It follows from Theorem~\ref{thm:Banach} that a sufficient condition for the convergence of Algorithms \ref{pnp-fbs} and \ref{red-pg} is that operators $\T_{\mathrm{PnP}}$ in \eqref{PnPop} and $\T_{\mathrm{RED}}$ in \eqref{REDop} are contractive. In fact, this is the case if the denoiser $\D$ is contractive and the loss function $f$ is convex.

\vspace{-3mm}
\begin{proposition}
\label{prop:contraction}
Let $f$ be convex and let $\D$ be contractive. Then the operator in \eqref{REDop} is contractive for all $\gamma >0$ and $L >1$. Moreover, if $f$ is $\beta$-smooth,  then operator \eqref{PnPop}  is contractive for all $0 < \gamma < 2/\beta$.
\end{proposition}
\vspace{-3mm}

\begin{proof}
If $\D$ is contractive and $L>1$, then it is not difficult to see that the second operator in \eqref{REDop} is contractive, being the convex combination of $\I$ and $\D$. On the other hand, the first operator in \eqref{REDop} is a proximal operator, which is nonexpansive \cite{bauschke2017convex}. Being the composition of a nonexpansive and a contractive operator, \eqref{REDop} is contractive.

On the other hand, if $f$ is convex and $\beta$-smooth, then from the Baillon-Haddad theorem \cite{bauschke2009baillon} we have that, for all $\x, \y \in \Re^n$,
\begin{equation*}
\beta\, \big(\nabla \! f(\x) -\nabla \! f(\y) \big)^\top\! (\x-\y) \geqslant \big\lVert\nabla \! f(\x) -\nabla \! f(\y) \big\rVert^2.
\end{equation*}
Using this, we can easily show that \eqref{gradstepop} is nonexpansive for all $0 < \gamma < 2/\beta$. Thus, \eqref{PnPop} is the composition of a nonexpansive and a contractive operator and is hence contractive.
\end{proof}

We arrive at the following important conclusion by combining Theorem~\ref{thm:Banach} and Proposition~\ref{prop:contraction}.

\vspace{-3mm}
\begin{corollary}
\label{corr:Dcontractive}
Suppose  $f$ is convex and $\D$ is contractive. 
\begin{enumerate}
\item Algorithm \ref{red-pg} converges globally and linearly if $\gamma >0$ and $L >1$.
\item Additionally, if $f$ is $\beta$-smooth, then Algorithm \ref{pnp-fbs} converges globally and linearly if $0 < \gamma < 2/\beta$.
\end{enumerate}
\end{corollary}
\vspace{-3mm}

\begin{proof}
If  $f$ is convex, $\gamma >0$ and $L >1$, and $\D$ is contractive, we know from Proposition~\ref{prop:contraction} that  \eqref{REDop} is a contraction. Similarly, if $0 < \gamma < 2/\beta$, then it follows from Proposition~\ref{prop:contraction} that  \eqref{PnPop}  is a contraction. Hence, we have from Theorem~\ref{thm:Banach} that Algorithms \ref{pnp-fbs} and \ref{red-pg} converges globally and linearly. 
\end{proof}

The convergence results in Corollary \ref{corr:Dcontractive} can be extended to averaged denoisers. This is important as we will later work with averaged denoisers that are not guaranteed to be contractive. We will rely on the following fixed-point convergence result for averaged operators.

\vspace{-3mm}
\begin{theorem}
\label{thm:Mann}
Let $\T$ be an averaged operator on $\R^n$ and assume that $\mathrm{Fix} \, \T \neq \emptyset$. The sequence $\{\x_k\}_{k \geqslant 0}$ generated as  $\x_{k+1} = \T(\x_k)$ converges globally to $\mathrm{Fix} \, \T$.
\end{theorem}
\vspace{-3mm}

The iteration $\x_{k+1} = \T(\x_k)$ is called the Mann iteration when $T$ is averaged, and Theorem~\ref{thm:Mann} is the Krasnosel’skii-Mann theorem \cite[Prop. 5.16]{bauschke2017convex}. %\cite{Mann1953,Kasnoselskii1955}

\vspace{-3mm}
\begin{remark}
Comparing Theorems \ref{thm:Banach} and  \ref{thm:Mann}, we note that while $\mathrm{Fix} \, \T$ is guaranteed to be nonempty when $\T$ is contractive, we need to assume this if $\T$ is averaged. Second, when $\T$ is a contraction, the sequence converges to the same fixed point regardless of the initialization $\x_0$, whereas the limit point will generally depend on $\x_0$ if $\T$ is averaged. Finally, the convergence in Theorem~\ref{thm:Mann} cannot be guaranteed to be linear.
\end{remark}
\vspace{-3mm}
%PnP-FBS and PnP-ADMM, where $\prox_{\rho g}$ is replaced by a denoising operator $\D$ \cite{Ryu2019_PnP_trained_conv,Sun2019_PnP_SGD,nair2021fixed}. We just need to ensure that $\D$ is contractive or averaged. Indeed, the proof of Theorem~\ref{convthm1} goes through in this case. More precisely, PnP-FBS converges  if $\D$ is $\theta$-averaged where $\theta \in (0,1)$, and PnP-ADMM converges if  $\D$ is $\theta$-averaged where $\theta \in (0,1/2]$ \cite{nair2021fixed}. 

If the denoiser $\D$ is nonexpansive or averaged, then we can make the following conclusion regarding the PnP and RED operators. 

\vspace{-3mm}
\begin{proposition}
\label{prop:averaged}
Let $f$ be convex and let $\D$ be nonexpansive. Then the operator in \eqref{REDop} is an averaged operator for all $\gamma >0$ and $L >1$. Moreover, if $f$ is $\beta$-smooth and $\D$ is averaged, then operator \eqref{PnPop} is an averaged operator for all $0 < \gamma < 1/\beta$.
\end{proposition}
\vspace{-3mm}

\begin{proof}
The reasoning is similar to that of Proposition~\ref{prop:contraction}. The second operator in \eqref{REDop} is $1/L$-averaged if $L>1$. Also, it is well-known that the proximal operator of a convex function is firmly nonexpansive \cite{bauschke2017convex}. Thus, being the composition of two averaged operators, we get that \eqref{REDop} is averaged \cite{bauschke2017convex}.

If $f$ is $\beta$-smooth and $0 < \gamma < 1/\beta$, we can show that \eqref{gradstepop} is firmly nonexpansive \cite{bauschke2017convex}. Again, being the composition of two averaged operators, \eqref{PnPop} must be averaged.
\end{proof}

Combining Theorem~\ref{thm:Mann} and Proposition~\ref{prop:averaged}, we can certify convergence of Algorithms \ref{pnp-fbs} and \ref{red-pg}.

\vspace{-3mm}
\begin{corollary}
\label{corr:Daveraged}
Suppose $f$ is convex and $\D$ is averaged. 
\begin{enumerate}
\item Assuming that $\mathrm{Fix} \, \T_{\mathrm{RED}} \neq \emptyset$, Algorithm \ref{red-pg} converges globally and for any $\gamma >0, \, L >1$.
\item Assuming that $\mathrm{Fix} \, \T_{\mathrm{PnP}} \neq \emptyset$ and $f$ is $\beta$-smooth, Algorithm \ref{pnp-fbs} converges globally for any $0 < \gamma < 1/\beta$.
\end{enumerate}
\end{corollary}
\vspace{-3mm}

\begin{proof}
If $f$ is convex, $\gamma >0$, $L >1$, and $\D$ is averaged (and hence nonexpansive), it follows from Proposition~\ref{prop:averaged} that \eqref{REDop} is an averaged operator. Thus, if $\mathrm{Fix} \, \T_{\mathrm{RED}} \neq \emptyset$, then we have from  Theorem~\ref{thm:Mann} that Algorithm \ref{red-pg} converges globally. 

On the other hand, if $f$ is convex and $\beta$-smooth, and $0 < \gamma < 1/\beta$, then it follows from Proposition~\ref{prop:averaged} that \eqref{PnPop} is an averaged operator. Hence, if $\mathrm{Fix} \, \T_{\mathrm{PnP}} \neq \emptyset$, then we have from Theorem~\ref{thm:Mann} that Algorithm \ref{pnp-fbs} converges globally.
\end{proof}

In summary, we were able to establish fixed-point convergence for Algorithms \ref{pnp-fbs} and \ref{red-pg} for contractive (Corollary \ref{corr:Dcontractive}) and averaged denoisers (Corollary \ref{corr:Daveraged}). We next show how such denoisers can be realized using deep unfolding.

\vspace{-3mm}
\begin{remark}
We can work out convergence results similar to Corollaries \ref{corr:Dcontractive} and \ref{corr:Daveraged} for the PnP and RED algorithms in \cite{romano2017little,reehorst2018regularization,sun2019block,cohen2021regularization}. 
\end{remark}
\vspace{-3mm}

\section{Unfolded Denoisers}
\label{proposed}

Motivated by the previous results, we wish to develop trained denoisers that are either contractive or averaged by construction and whose regularization capacity is comparable to state-of-the-art CNN denoisers. As explained in Section \ref{motivation}, learning a nonexpansive CNN denoiser presents several challenges. This prompted us to consider a different solution to the problem, namely, using unfolded networks. More specifically, we construct the denoiser in three steps.

\begin{enumerate}
\item We consider the classical variational problem of wavelet denoising and unfold it using FBS and DRS. This is done at a patch level. The patch denoiser is contractive (for FBS) or averaged (for DRS) by construction.
\item We train the patch denoiser using supervised learning, where the regularization parameters (thresholds) of wavelet denoising and the internal (step size and penalty) parameters of FBS and DRS are optimized. 
\item We aggregate the patch denoisers to derive an image denoiser, which is proven to inherit the contractive or averaged property of the original patch denoiser.
\end{enumerate}

\subsection{Wavelet denoising}

Let $\b \in \Re^p$ represent a noisy image patch, where we have stacked the pixels of a $(k \times k)$ image patch into a vector of length $p=k^2$. Let $\W$ be an orthogonal wavelet transform. Consider the wavelet denoising problem
\begin{equation*}
\underset{\x \in \Re^p}{\min} \ \   \frac{1}{2} \lVert \x - \b \rVert^2 + \lambda \sum_{i=1}^p  \, \lvert (\W \x)_i \rvert \qquad (\lambda >0), 
\end{equation*}
where $\lambda$ is the regularization parameter. The above problem has the closed-form solution $\W^\top \phi_{\lambda}  (\W \b)$, 
where the soft-threshold function $$\phi_{\lambda}(\theta) = \mathrm{sign}(\theta)\, \max (0, \lvert \theta \rvert - \lambda)$$ is applied componentwise on $\W \b$ and $\lambda$ acts as a threshold \cite{chambolle1998nonlinear}.

We consider a slightly different model with a different threshold for each wavelet coefficient. Namely, we consider the optimization problem
\begin{equation}
\label{wt}
\underset{\x \in \Re^p}{\min} \ \   \frac{1}{2} \lVert \x - \b \rVert^2 + \sum_{i=1}^p \lambda_i \, \lvert (\W \x)_i \rvert \quad (\lambda_i >0).
\end{equation}
This also has a closed-form solution given by 
\begin{equation}
\label{proxwav}
\phi_{\tri}(\b) = \W^\top  \S_{\tri} (\W\b),
\end{equation}
where $\tri = (\lambda_1,\ldots,\lambda_p)$ are the thresholds and the thresholding operator $ \S_{\tri}: \Re^p \to \Re^p$ is defined as  
\begin{equation*}
\forall \, \boldsymbol{\theta} \in \Re^p, \, i \in [1,p]: \quad \big( \S_{\tri}(\boldsymbol{\theta}) \big)_i = \phi_{\lambda_i}(\theta_i).
\end{equation*}

\subsection{FBS denoiser}

The denoising operator $\phi_{\tri}$ is nonexpansive, being the composition of nonexpansive and unitary operators. However, it is easy to see that $\phi_{\tri}$ is not contractive. To identify a contractive (resp.~averaged) operator, we propose to solve the denoising problem \eqref{wt} in an iterative fashion using FBS (resp.~DRS). The proposed denoising network is obtained by chaining the update operator in FBS and DRS.

As a first step, we apply FBS to the optimization problem \eqref{wt}. Note that,  a stepsize $\rho$, we can write the FBS update as $$\x_{k+1} = \U(\x_k; \rho, \tri),$$ where we define the FBS block $\U$ to be
\begin{equation}
\label{FBSlayer}
\U(\x; \rho, \tri) = \W^\top \S_{\tri} \left (\W \big((1 - \rho)\x + \rho \b \big) \right). 
\end{equation}

\begin{figure*}
\centering
\subfloat[FBS block]{\includegraphics[width=1.0\linewidth]{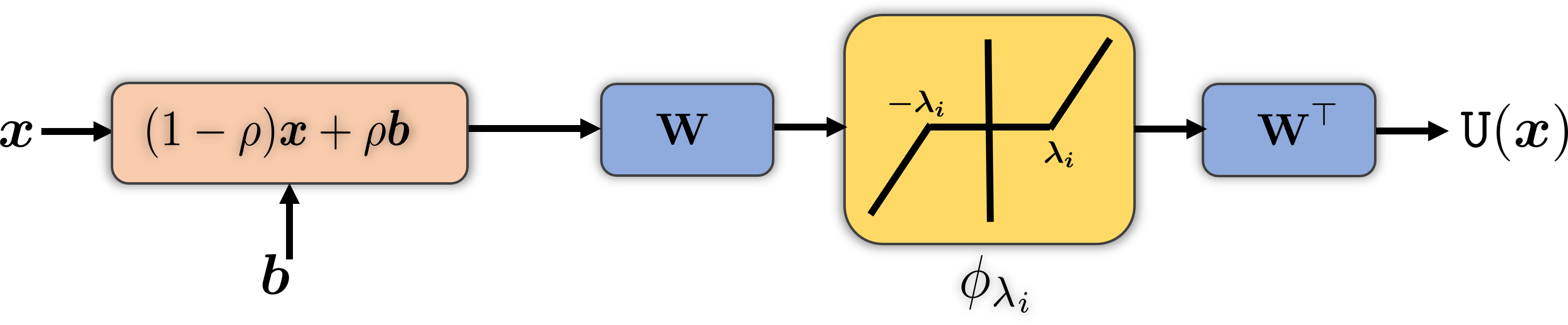}} \\
\subfloat[DRS block]{\includegraphics[width=1.0\linewidth]{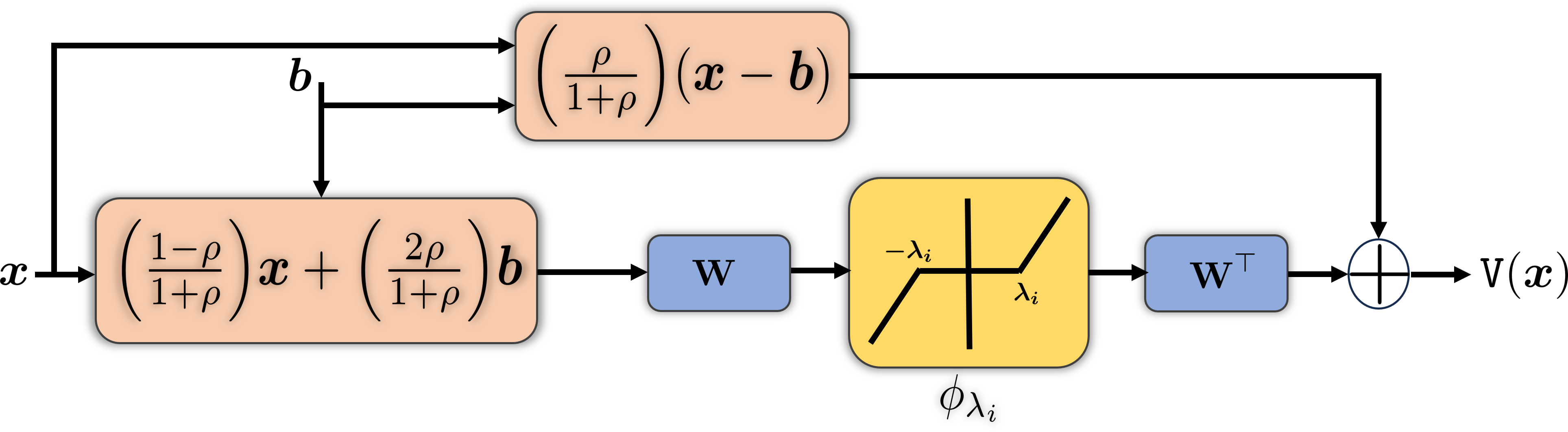}}
\caption{Building blocks obtained by unfolding FBS and DRS for wavelet denoising.%, given a noisy patch $\b$. 
The proposed denoisers are obtained by chaining these blocks.}
\label{buildingblocks}
\end{figure*}

The FBS block is depicted in Fig.~\ref{buildingblocks}. Functionally, $\U$ is an inversion step (gradient descent on the loss) followed by wavelet regularization. %As is well known \cite{bauschke2017convex}, for any $\x_0 \in \Re^p$ and $\rho \in (0,1]$,  the iterates $\x_{k+1} = \U(\x_k; \rho, \tri)$ converge to the solution  of the original problem \eqref{wt}. 

Although \eqref{wt} has a closed-form solution, we are interested in the iterative solution of \eqref{wt} because of the following property.

\vspace{-3mm}

\begin{proposition}
\label{prop:ctr}
For all $\rho \in (0,1]$ and $\tri >0$, the map $\x \mapsto \U(\x; \rho, \tri)$ is contractive.
\end{proposition}
\vspace{-3mm}

Indeed, since $\W$ is an orthogonal transform and the operator $\S_{\tri}$ in \eqref{FBSlayer} is nonexpansive (the condition $\tri >0$ is required to ensure that \eqref{proxwav} is well-defined), we have for any $\x, \x' \in \R^n$,
\begin{align*}
\lVert \U(\x; \rho, \tri)  - \U(\x'; \rho, \tri)  \rVert \leqslant  (1-\rho) \lVert \x - \x' \rVert.
\end{align*}
Since $1-\rho \in [0,1)$, we can conclude that  the map $\x \mapsto \U(\x; \rho, \tri)$ is contractive.

Following Proposition~\ref{prop:ctr}, we chain $\U$ into a network. The precise definition is as follows.

\vspace{-3mm}
\begin{definition}[FBS denoiser]
For a fixed number of layers $L \geqslant 1$, we define $\D_{\mathrm{FBS}}: \Re^p \to \Re^p$ to be
\begin{equation}
\label{fbsdenoiser}
\D_{\mathrm{FBS}} = \U^{(L)} \circ \U^{(L-1)} \circ \cdots \circ \U^{(1)},
\end{equation}
where each layer has the form $\U^{(\ell)} = \U(\cdot \ ;\rho^{(\ell)}, \tri^{(\ell)})$.
\end{definition}
\vspace{-3mm}

Since the composition of contractive operators is again contractive, the next result follows immediately from Proposition~\ref{prop:ctr}. 

\vspace{-3mm}
\begin{proposition}
\label{Dfbs-contr}
For all $\rho^{(\ell)} \in (0,1]$ and $\tri^{(\ell)} >0$, $\D_{\mathrm{FBS}}$ is contractive.
\end{proposition}

\vspace{-3mm}

The point is that \eqref{fbsdenoiser} is contractive as long as $\tri^{(\ell)}$ and $\rho^{(\ell)}$ are in the interval stipulated in Proposition~\ref{Dfbs-contr}. We can thus tune the parameters over these intervals to optimize the denoising capacity while preserving the contractive nature.
%without having to enforce structural constraints during training. 

\vspace{-3mm}
\begin{remark}
Unlike LISTA \cite{gregor2010learning} or DualFB \cite{repetti2022dual}, we use a fixed transform in each layer. This is because we wish to retain the proximal structure in \eqref{FBSlayer}, and orthogonality plays a crucial role in this regard. %It is known to be challenging to train a CNN under orthogonality constraint \cite{kiani2022projunn}.
\end{remark}
\vspace{-3mm}

\subsection{DRS denoiser}

Note that we can apply Peaceman-Rachford \cite{bauschke2017convex} or DRS to problem  \eqref{wt}. In fact, we will show that the corresponding update operator is firmly nonexpansive for DRS. Thus, the resulting denoiser obtained by chaining this operator is averaged, and we can guarantee convergence of Algorithms \ref{pnp-fbs} and \eqref{red-pg} (Corollary \ref{corr:Daveraged}). 

\vspace{-2mm}

\begin{remark}
It is well-known that DRS converges in fewer iterations than FBS \cite{parikh2014proximal}. Since each layer of the unfolded denoiser corresponds to an iteration, we expect to perform better denoising with DRS for the same number of layers. Indeed, we will show in Section \ref{Experiments} that the DRS denoiser gives better results than the FBS denoiser in some experiments. 
\end{remark}

\vspace{-2mm}

To arrive at the DRS update for \eqref{wt}, we identify \eqref{wt} with \eqref{mainopt} by setting  
\begin{equation*}
f(\x) = \frac{1}{2} \lVert \x-\b \rVert^2, \quad g(\x)=  \sum_{i=1}^p \lambda_i \, \lvert (\W \x)_i \rvert.
\end{equation*}
It follows from \eqref{drs} that we can write the DRS update as written as $$\x_{k+1} = \V(\x_k; \rho, \tri),$$ where the DRS block $\V$ is defined as 
\begin{align}
\label{eq:Vblock}
&\V(\x; \rho, \tri) =\frac{1}{2} \x \nonumber  \\  
&+ \frac{1}{2}   \big(2\prox_{\rho g} - \I  \big) \big(2\prox_{\rho f} - \I\big)(\x),
\end{align}
and $\rho >0$ is the penalty parameter of DRS. 

\vspace{-3mm}
\begin{remark}
We can simplify the expression for the DRS block \eqref{eq:Vblock}. Indeed, a direct calculation gives
\begin{equation*}
\prox_{\rho f}(\x)= \frac{1}{1+\rho}( \x + \rho \b).
\end{equation*}
On the other hand, if we absorb $\rho$ in $\tri$, we get $\prox_{\rho g}=\phi_{\tri}$. Thus, we can express \eqref{eq:Vblock} as
\begin{align}
\label{drsblock}
\V(\x; \rho, \tri) = &\frac{\rho}{1+\rho} (\x-\b)  & \nonumber\\
&+\phi_{\tri}\left(\frac{1-\rho}{1+\rho} \x + \frac{2 \rho}{1+\rho} \b \right).
\end{align}
\end{remark}
\vspace{-3mm}

Having defined the update operator $\V$, we can now establish its averaged property. 

\vspace{-3mm}
\begin{proposition}
\label{prop:fNE}
For all $\rho>0$ and $\tri >0$, the map $\x \mapsto \V(\x; \rho, \tri)$ is firmly nonexpansive.
\end{proposition}
\vspace{-3mm}

Indeed, we know that the proximal operator of a convex function is firmly nonexpansive \cite{bauschke2017convex}. Thus, the operators 
$2\prox_{\rho g} - \I$ and $2\prox_{\rho f} - \I$ are nonexpansive. Thus, if we denote their composition by $\N$, then $\N$ must be nonexpansive. Since  $\V=(1/2) (\I + \N)$,  $\V$ must be firmly nonexpansive.

We will now use Proposition~\ref{prop:fNE} to construct an averaged network. As with the FBS block, the trainable parameters in \eqref{drsblock} are $\rho$ and $\tri$, while $\b$ and $\W$ are fixed. The DRS block is depicted in Fig.~\ref{buildingblocks}. The DRS denoiser is obtained by chaining $\V$ into a network.

\vspace{-3mm}
\begin{definition}[DRS denoiser]
For a fixed number of layers $L \geqslant 1$, we define $\D_{\mathrm{DRS}}:\Re^p \to \Re^p$ to be
\begin{equation}
\label{drsdenoiser}
\D_{\mathrm{DRS}} = \V^{(L)} \circ \V^{(L-1)} \circ \cdots \circ \V^{(1)},
\end{equation}
where each layer is of the form $\V^{(\ell)} = \V(\cdot \ ;\rho^{(\ell)}, \tri^{(\ell)})$.
\end{definition}
\vspace{-3mm}

As a consequence of Proposition~\ref{prop:fNE}, we have the following structure on the denoiser.
\vspace{-3mm}
\begin{proposition}
\label{Ddrs-avgd}
For all $\rho^{(\ell)} >0$ and $\tri^{(\ell)} >0$, $\D_{\mathrm{DRS}}$ is $L/(L+1)$-averaged.
\end{proposition}
\vspace{-3mm}

Indeed, notice from Proposition~\ref{prop:fNE} that $\D_{\mathrm{DRS}}$ is the composition of $L$ firmly nonexpansive operators. Now,  if we compose $L$ number of $\theta$-averaged operators, then it can be shown that the resulting operator is $L\theta/(1+(L-1)\theta)$-averaged (e.g., see \cite{bauschke2017convex}). Setting $\theta=1/2$, we arrive at Proposition~\ref{Ddrs-avgd}.

\begin{figure*}[!htp]
\captionsetup[subfloat]{labelformat=empty}
\centering
  \subfloat[\tiny barbara]{\includegraphics[width=0.0785\linewidth]{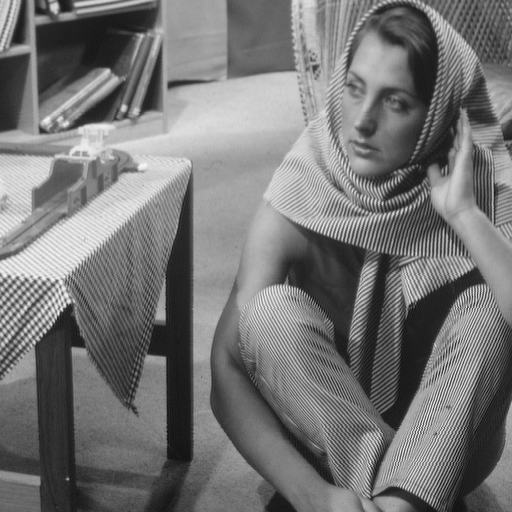}} \hspace{0.01mm}
  \subfloat[\tiny ship]{\includegraphics[width=0.0785\linewidth]{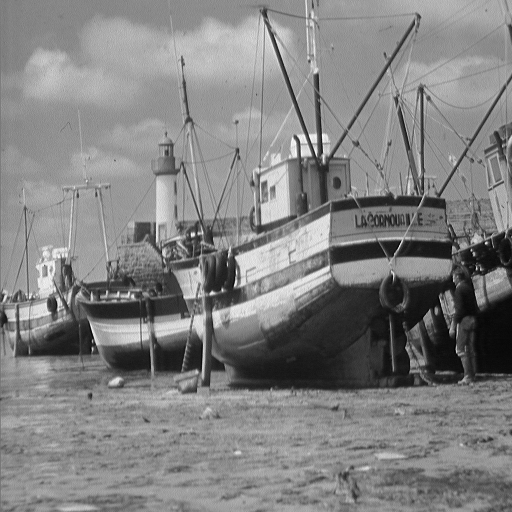}} \hspace{0.01mm}
  \subfloat[\tiny cameraman]{\includegraphics[width=0.0785\linewidth]{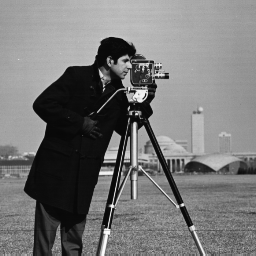}} \hspace{0.01mm}
  \subfloat[\tiny couple]{\includegraphics[width=0.0785\linewidth]{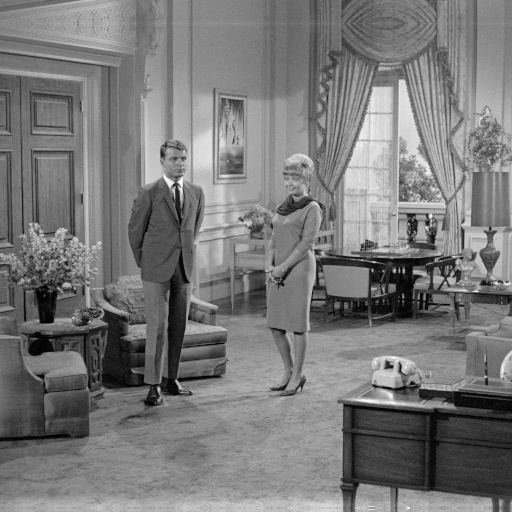}} \hspace{0.01mm}  
  \subfloat[\tiny fingerprint]{\includegraphics[width=0.0785\linewidth]{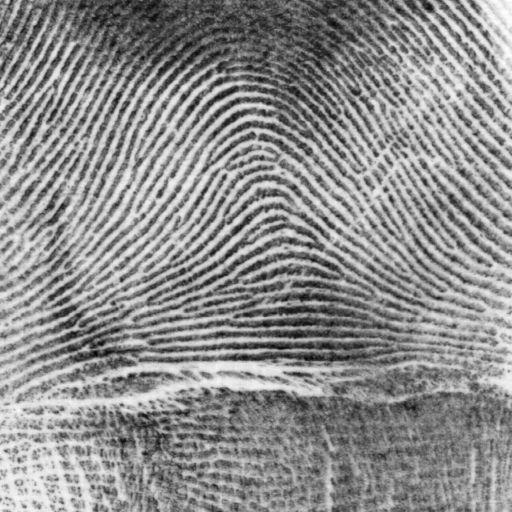}} \hspace{0.01mm}
  \subfloat[\tiny hill]{\includegraphics[width=0.0785\linewidth]{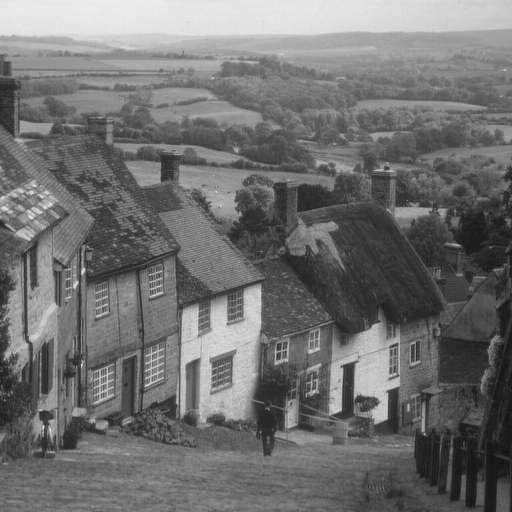}} \hspace{0.01mm}
  \subfloat[\tiny house]{\includegraphics[width=0.0785\linewidth]{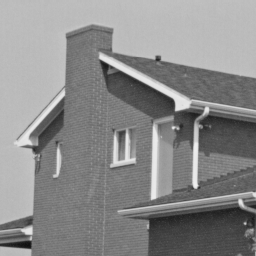}} \hspace{0.01mm}
  \subfloat[\tiny lena]{\includegraphics[width=0.0785\linewidth]{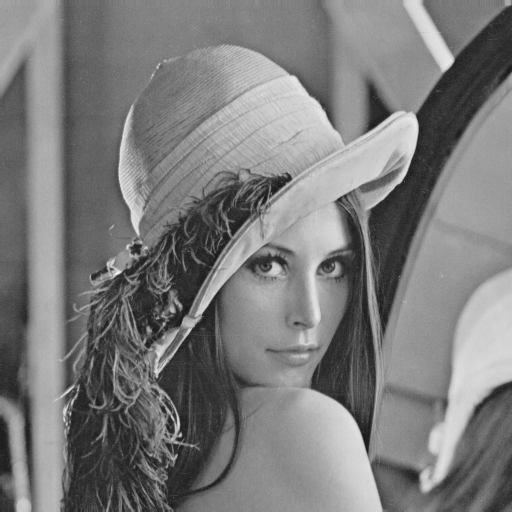}} \hspace{0.01mm}  
  \subfloat[\tiny man]{\includegraphics[width=0.0785\linewidth]{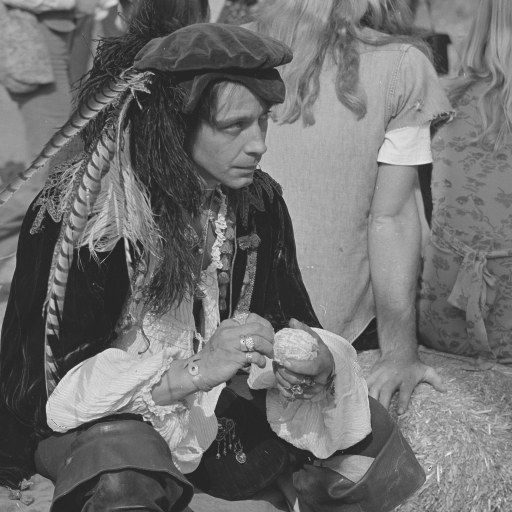}} \hspace{0.01mm}
  \subfloat[\tiny montage]{\includegraphics[width=0.0785\linewidth]{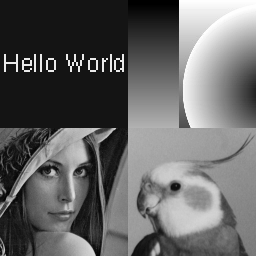}} \hspace{0.01mm}
  \subfloat[\tiny peppers]{\includegraphics[width=0.0785\linewidth]{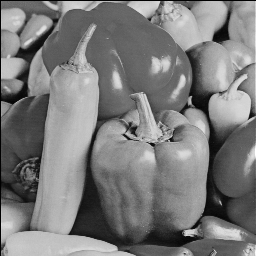}} \hspace{0.01mm}
  \subfloat[\tiny starfish]{\includegraphics[width=0.0785\linewidth]{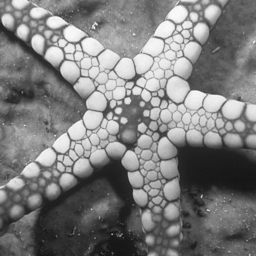}} 
   \caption{Grayscale images from the Set12 database. We will identify these images by their corresponding names in Section \ref{Experiments}.} 
\label{Inputimages}
\end{figure*}

\subsection{Image denoiser}
\label{sec:imgdenoiser}

Recall that \eqref{fbsdenoiser} and \eqref{drsdenoiser} are patch denoisers. We apply them independently on patches extracted from a noisy image and then average the denoised patches to obtain the final image \cite{zhang2017image,dabov2007image}. The overall operation defines an image-to-image denoising operator $\D$, which is used in Algorithms \ref{pnp-fbs} and \ref{red-pg}. In view of Corollary \ref{corr:Dcontractive} and \ref{corr:Daveraged}, we wish to ensure that $\D$ is averaged or contractive. First, we give a precise definition of the denoiser and then prove that it has the desired property.

Let $\Omega=[0,q-1]^2$ denote the image domain and $\X: \Omega \to \Re$ be the input (noisy) image; we also view $\X$ as a vector in $\Re^n$ where $n=q^2$. 
%Let $\x \in \Re^n$ denote the vectorized image, where $n=q^2$ is the number of pixels. 
We assume periodic boundary condition, i.e., for $\i = (i_1,i_2) \in \Omega$ and $\btau =(\tau_1,\tau_2) \in \mathbb{Z}^2$, we define 
\begin{equation*}
\i - \btau = \big( (i_1 - \tau_1) \ \mathrm{mod} \ q, \, (i_2 - \tau_2) \ \mathrm{mod} \ q\big).
\end{equation*}
Without loss of generality, we assume that $q$ is a multiple of the patch size $k$; we can simply pad the image to satisfy this condition.

Let the patch size be $k \times k$. For $\i \in \Omega$, we define the patch operator
\begin{align}
\label{patchop}
&\Pcal_{\i}: \Re^n \to \Re^{k^2}, \nonumber \\
\Pcal_{\i}(\X)=\mathrm{vec}& \Big( \big\{\X(\i + \btau):  \btau \in [0,k-1]^2\big\} \Big),
\end{align}
where $\mathrm{vec}(\cdot)$ denotes the vectorized representation of the pixels in a $(k \times k)$ window around $\i$ (in some fixed order).

We extract patches from $\X$ with stride $s$, where we assume that $s$ divides $k$. 
Thus, the starting coordinates of the patches are $\J = [0,(q_s-1)]_s^2$, where $q_s =  q/s$. 
%Recall from Section \ref{notationsec} that $[0,(q_s-1)]_s$ denotes all the integers in $[0,(q_s-1)s]$ which are multiples of $s$. 
In other words, the patches are $$\left\{\Pcal_{\j}(\X): \, \j \in \J\right\}.$$

\vspace{-3mm}
\begin{definition} Given a patch denoiser $\D_{\mathrm{patch}}$, we define the image denoiser $\D: \Re^n \to \Re^n$ to be
\begin{equation}
\label{imagedenoiser} 
{\D}(\X) = \frac{1}{(k/s)^2} \sum_{\j \in \J} \Pcal_{\j}^* \big(\D_{\mathrm{patch}} (\Pcal_{\j}(\X)) \big) ,
\end{equation}
where $\Pcal_{\j}^*$ is the adjoint of $\Pcal_{\j}$. 
\end{definition}
\vspace{-3mm}

The factor $(k/s)^2$ is the number of patches containing a given pixel; the assumption that $s$ divides $k$ is essential to ensure this is the same for every pixel.

\vspace{-3mm}
\begin{remark}
The adjoint in \eqref{imagedenoiser} is 
is w.r.t. the standard Euclidean inner-product (see Appendix \ref{proof:imagedenoiser} for details). In particular, it is a linear map from $\Re^{k^2}$ to $\Re^n$ such that for any patch $\x \in \Re^{k^2}$, $\Y=\Pcal^*_{\j}(\x)$ is an image, where  
\begin{equation*}
\Pcal_{j}(\Y) = \x,
\end{equation*}
and the rest of the pixels in $\Y$ are zero. 
\end{remark}
\vspace{-3mm}
%We can show this as follows, if $r$ is the number of patches a pixel belongs to, then $rq^2 = k^2 q_s^2$ by double counting and hence $r=k_s^2$. In \eqref{imagedenoiser}, $\lvert \J \rvert = q_s^2$ is the total number of extracted patches.  
%When ISTA blocks are used in $\D_p$, we denote the image denoiser in \eqref{imagedenoiser} as ${\D}_{\ISTAop}$ and when ADMM blocks are used, we denote \eqref{imagedenoiser} as ${\D}_{\ADMMop}$. We have the following result for ${\D}_{\ISTAop}$ and ${\D}_{\ADMMop}$. 

Henceforth, we will assume that the patch denoiser is either $\D_{\FBSop}$ 
 in \eqref{fbsdenoiser} or $\D_{\DRSop}$ in \eqref{drsdenoiser}. We now state the main result of this section.

\vspace{-3mm}
\begin{proposition}
\label{finalprop}
The image denoiser \eqref{imagedenoiser} derived from \eqref{fbsdenoiser} is contractive, while the denoiser derived from  \eqref{drsdenoiser} is averaged.
\end{proposition}
\vspace{-3mm}

The proof is given in Appendix \ref{proof:imagedenoiser}. The core idea is to express 
 the original system of overlapping patches in terms of non-overlapping patches. This allows us to use some form of orthogonality in the calculations.

%Since contractive and averaged operators are nonexpansive, the proposed denoiser can be used to guarantee convergence for any iterative algorithm that requires the denoiser to be contractive, averaged, or nonexpansive. In particular, we record the following important implication of the results so far. 

%\begin{corollary}
%\label{maincorollary}
%Suppose the loss term $f$ is convex in \eqref{mainopt}. Then if we plug $\D_{\mathrm{FBS}}$ or $\D_{\mathrm{ADMM}}$ into PnP-FBS \cite{nair2021fixed}, RED-PG \cite{reehorst2018regularization} and algorithms such as \cite{Sun2019_PnP_SGD,sun2019block,sun2021scalable,cohen2021regularization}, we can guarantee iterate convergence of the corresponding algorithm.
%\end{corollary}

\subsection{Training}

We trained \eqref{fbsdenoiser}  and \eqref{drsdenoiser} on patches extracted from images of the BSD500 dataset. Let $\x_1,\ldots,\x_N \in \R^p$ be the clean patches (normalized to  $[0,1]$ intensity), where $p=k^2$ and $N$ is the size of traning data. The noisy patches are generated as follows:
\begin{equation*}
{\z}_i = \x_i + \sigma \boldsymbol{\eta},
\end{equation*}
where $\boldsymbol{\eta} \sim \mathcal{N}(\boldsymbol{0}, \I)$ and $\sigma$ is the noise level. Thus, the training examples are $({\z}_1, \x_1), \ldots,  ({\z}_N, \x_N)$.

Let us use $\Theta$ to denote the parameters in  \eqref{fbsdenoiser}  and \eqref{drsdenoiser}. We will write the patch denoiser as $\D_\text{patch}(\cdot \,;\Theta)$ to account for the dependence on $\Theta$.
%\begin{equation*}
%\Theta = \Big\{\rho^{(\ell)}, {\lambda_j}^{(\ell)}: \ \ell = 1, \ldots, L,\, j = 1, \ldots ,p \Big \}.
%\end{equation*}
The training problem is to optimize $\Theta$ such that $\D_\text{patch}({\z}_i;\Theta) \approx \x_i$ for all $i=1,\ldots,N$.
Moreover, in view of Proposition~\ref{Dfbs-contr}, we consider the following optimization problem for training \eqref{fbsdenoiser}:

\begin{equation}
\label{trainDfbs}
\begin{aligned}
\underset{\Theta}{\min} \quad &  \sum_{i=1}^N \ {\lVert  \D_\text{FBS}({\z}_i; \Theta) - \x_i \rVert}^2 \\
\textrm{s.t.} \quad & \rho^{(1)},\ldots, \rho^{(L)} \in [\rho_0, 1], \\
& {\lambda^{(1)}_j},\ldots, {\lambda^{(L)}_j}\in [\lambda_0, \infty) \quad 
\big( j \in [1, p] \big).
\end{aligned}
\end{equation}
The lower bounds $\rho_0 \in (0,1)$ and  $\lambda_0 >0$ are used to ensure that the constraint set in \eqref{trainDfbs} is closed; this will be useful when we project $\Theta$ onto the intervals.  

On the other hand, in view of Proposition~\ref{Ddrs-avgd}, we consider the following optimization for training \eqref{drsdenoiser}:
\begin{equation}
\label{trainDdrs}
\begin{aligned}
\underset{\Theta}{\min} \quad & \sum_{i=1}^N \ {\lVert  \D_\text{DRS}({\z}_i; \Theta) - \x_i \rVert}^2 \\
\textrm{s.t.} \quad & \rho^{(1)},\ldots, \rho^{(L)}  \in [\rho_0, \infty), \\
& {\lambda^{(1)}_j},\ldots, {\lambda^{(L)}_j}\in [\lambda_0, \infty) \quad \big( j \in [1, p] \big).
\end{aligned}
\end{equation}
%where  $\rho_0 \in (0,1)$ and  $\lambda_0 >0$.

\begin{table*}[!htp]
\caption{\textbf{Denoising} performance (PSNR/SSIM averaged) on the BSD68 dataset at different noise levels $\sigma$.}
\centering
\renewcommand{\arraystretch}{1.2}
%\resizebox{\columnwidth}{!}{ 
\begin{tabular}{ c|c c c c c } 
\hline
  &  BM3D &  DnCNN &  N-CNN &  $\D_{\mathrm{FBS}}$ &  $\D_{\mathrm{DRS}}$ \\
  \hline
 $\sigma = 5$ &  $37.52$/$0.9668$ &  $38.01$/$0.9696$  &  $33.35$/$0.9212$ &  $36.54$/$0.9480$ & $36.62$/$0.9506$\\ 
 $\sigma = 20$ &  $29.65$/$0.8799$ &  $30.26$/$0.8948$ & $27.99$/$0.8017$ &  $30.01$/$0.8344$ & $29.99$/$0.8346$\\
 $\sigma = 35$ & $27.10$/$0.7531$ &  $27.69$/$0.7809$ & $25.17$/$0.6252$ &  $26.18$/$0.6988$ &  $26.17$/$0.6952$\\
\hline
\end{tabular}%}
\label{Denoisingtab}
\end{table*}

\section{Experiments}
\label{Experiments}

For the unfolded network, we used a $10$-fold cascade of Daubechies, Haar, and Symlet orthogonal wavelets in succession (the number of layers is $L=30$). We used a fixed patch size $k=64$. We solved \eqref{trainDfbs} and \eqref{trainDdrs} using Adam optimizer \cite{kingma2014adam} with mini-batches of size $128$ and $5$ training epochs. After the gradient step, performed using automatic differentiation, we enforced the constraints in \eqref{trainDfbs} and \eqref{trainDdrs} by projecting $\Theta$ onto the respective intervals in \eqref{trainDfbs} and \eqref{trainDdrs}, essentially implementing a form of projected gradient descent. 
The lower bounds for trainable parameters in the optimization problems \eqref{trainDfbs} and \eqref{trainDdrs} were fixed to $\rho_0 = \lambda_0 = 0.0001$. We trained separate denoisers over the noise range $5\mbox{-}50$ in steps of $5$. The trained models ($500$ KB for each noise level) are lightweight compared to DnCNN ($2.3$ MB). 

The objective of the experiments in Sections \ref{subsec:denoise}, \ref{subsec:deblur}, and \ref{subsec:SR} 
is to analyze the denoising performance and the reconstruction accuracy of our denoisers (for deblurring and superresolution) and compare them with existing denoisers. Moreover, we present numerical evidence supporting the claimed contractive (averaged) property of the denoisers and the associated convergence guarantees for PnP and RED (Corollaries \ref{corr:Dcontractive} and \ref{corr:Daveraged}). We also present numerical evidence showing that state-of-the-art denoisers such as BM3D and DnCNN do not meet these guarantees.

\subsection{Denoising} 
\label{subsec:denoise}

%We analyze the proposed denoisers and compare their reconstruction accuracy and convergence with existing denoisers. 
%While analyzing the denoisers, we empirically demonstrate the nonexpansivity property of denoisers and provide numerical evidence for Corollary \ref{corr:Daveraged}.  
%We also present numerical results for image deblurring and superresolution by plugging our denoiser into PnP-FBS and RED-PG \cite{reehorst2018regularization}. 
%We show that our results on the Set 12 dataset are competitive with those obtained using the best denoisers. Importantly, as noted in Corollaries \ref{corr:Dcontractive} and \ref{corr:Daveraged}, iterate convergence for PnP-FBS and RED-PG is guaranteed for $\D_{\mathrm{FBS}}$ and $\D_{\mathrm{ADMM}}$. 

In Table \ref{Denoisingtab}, we compare the denoising performance of our denoisers with BM3D, DnCNN, and N-CNN on the Set12 dataset (see Fig.~\ref{Inputimages}). Our denoisers are off by PSNR of $0.25\mbox{-}1.5$ dB and SSIM of $0.02\mbox{-}0.08$ from the best denoiser DnCNN. However, they consistently outperform N-CNN across all noise levels. 
Recall that the purpose of our construction was to guarantee convergence of PnP and RED. In pursuit of this, we had to make design choices that resulted in a tradeoff in denoising performance. Importantly, we will see that the reconstruction capability of our denoiser is competitive with the above denoisers.
This is not surprising since there is no concrete relation between the denoising performance and the restoration capacity of a denoiser. For instance, it has been previously shown that BM3D can yield superior restoration results compared to DnCNN in specific applications \cite{Ryu2019_PnP_trained_conv}, even though DnCNN generally exhibits better denoising capabilities than BM3D. This suggests that the denoising gap is compensated in PnP and RED because the denoiser is applied in every iteration. However, this is just an intuitive guess, and we do not have a precise mathematical explanation.

\begin{figure*}[!htp]
\centering
\subfloat{\includegraphics[width=0.32\linewidth]{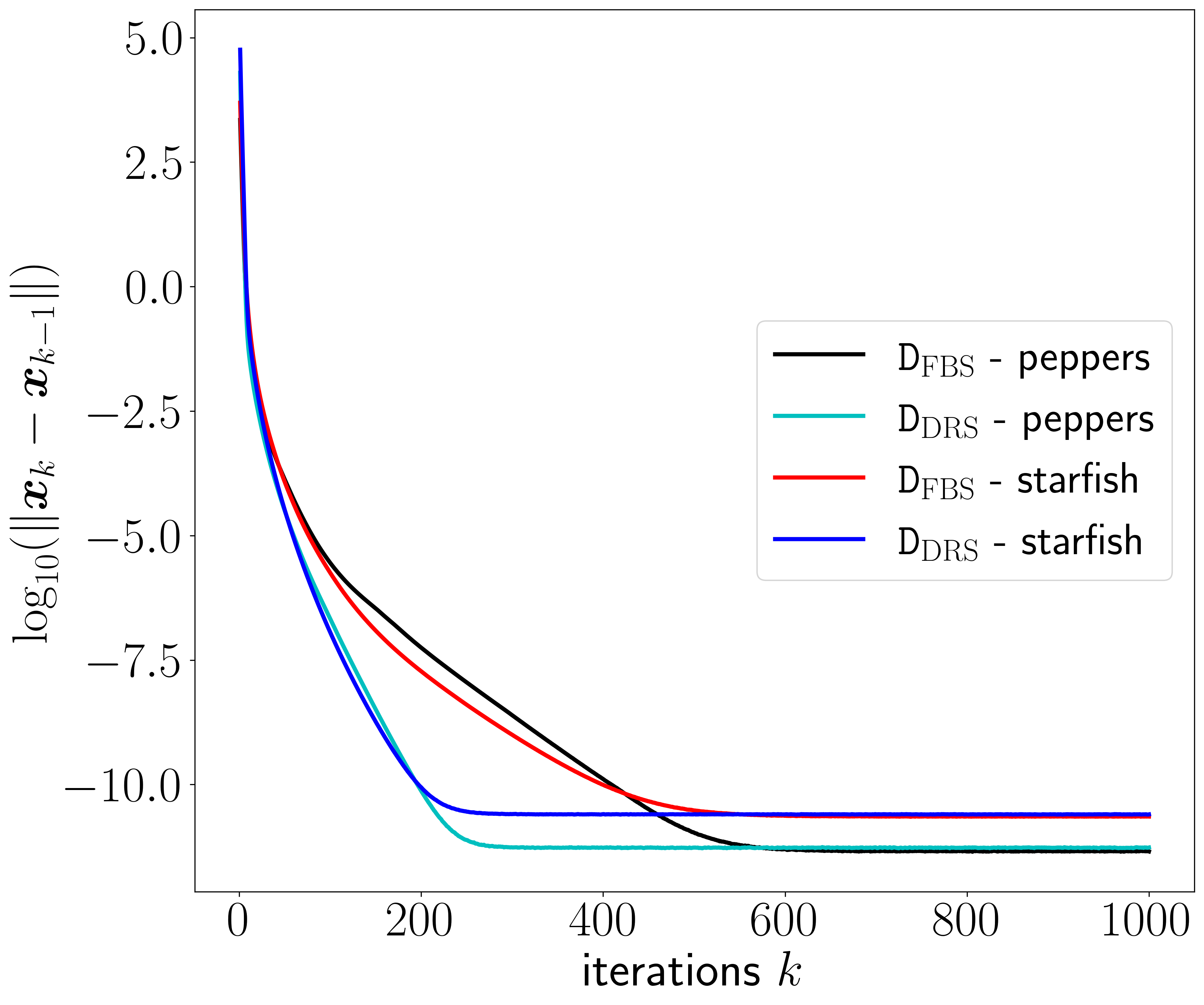}} \hspace{0.05mm}
\subfloat{\includegraphics[width=0.32\linewidth]{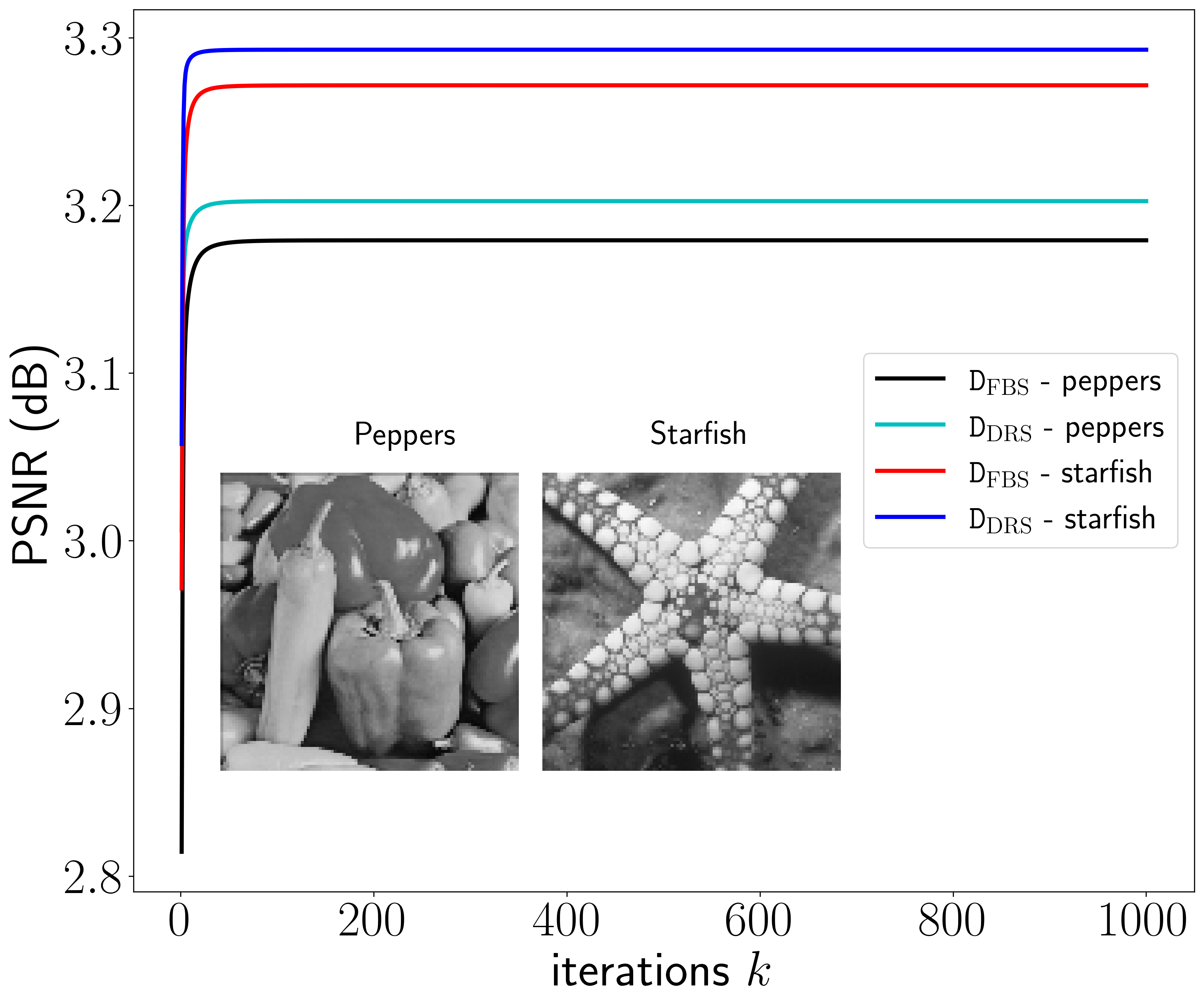}} \hspace{0.05mm}
\subfloat{\includegraphics[width=0.32\linewidth]{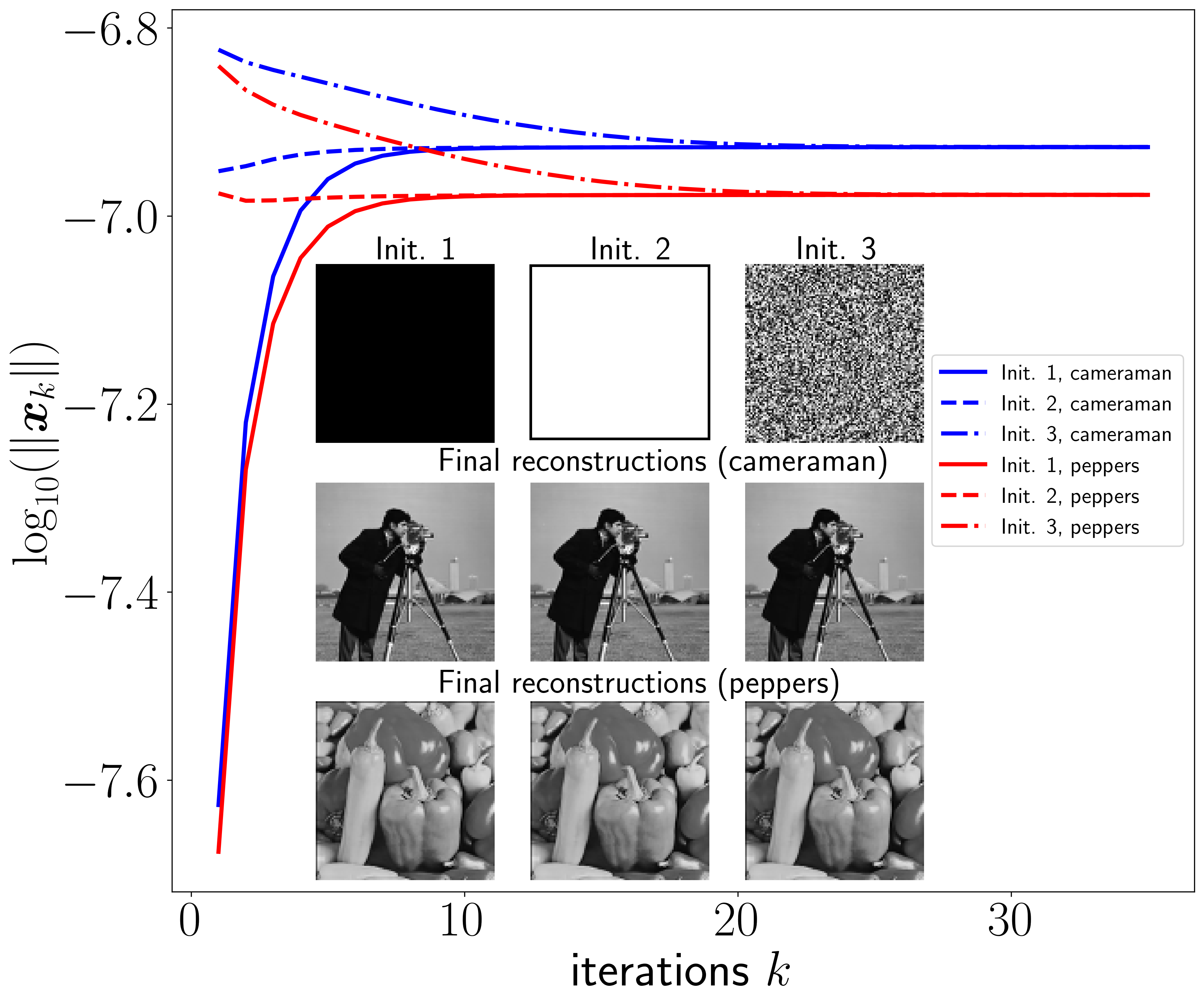}}
\caption{Analysis of the proposed denoisers. Denoisers $\mathrm{D}_{\FBSop}$ and $\mathrm{D}_{\DRSop}$ are plugged into PnP-FBS, which is used for image deblurring. We plot the difference between successive iterates and PSNR with iterations. We notice that the difference between iterates is strictly decreasing thanks to the averaged property, and the PSNR stabilizes in $50$-$100$ iterations. In (c), we validate that PnP-FBS converges to a unique fixed point with $\mathrm{D}_{\FBSop}$, thanks to the contraction property of operator $\T_{\mathrm{PnP}}$.
The experiment is performed with diverse initializations, namely, with $\x_0$ as all-zeros, all-ones, and random. 
The iterations stabilized in about $35$ iterations. The MSE between the reconstructions obtained using different initializations is of the order $1\mathrm{e}\mbox{-}8$.}
\label{propplots}
\end{figure*}

\begin{table*}[!htp]
\caption{Comparison of the residual $\|\x_k - \x_{k-1}\|$ and PSNR with DnCNN for a deblurring experiment. (see text for more details)}
\centering
\renewcommand{\arraystretch}{1.2}

%\resizebox{.45\textwidth}{!}{
\begin{tabular}{c| c| c c c c c }
\hline
\multirow{2}{*}{denoiser} & & \multicolumn{5}{c}{number of iterations} \\ 
& &  $10$ & $25$ & $50$ & $75$ & $100$  \\
\hline
\multirow{2}{*}{DnCNN} & $\|\x_k - \x_{k-1}\|$ & $146.00$ & $146.58$ & $234.55$ & $2300$ & $2207248$ \\
& PSNR & $16.54$ & $14.79$ & $-4.09$ & $-45.14$ & $-84.68$ \\
\hline
Proposed & $\|\x_k - \x_{k-1}\|$ &  $74.06$ & $30.65$ & $7.67$ & $1.99$ & $0.52$  \\ 
($\D_{\mathrm{FBS}}$) & PSNR & $20.42$ & $22.96$ & $23.69$ & $23.74$ & $23.74$ \\
\hline
Proposed & $\|\x_k - \x_{k-1}\|$ & $74.09$ & $30.67$ & $7.68$ & $1.99$ & $0.55$ \\ 
({$\D_{\mathrm{DRS}}$}) &  PSNR & $20.43$ & $23.00$ & $23.74$ & $23.79$ & $23.79$ \\
\hline
\end{tabular}%}
\label{counterexample1}
\end{table*}

We conducted a comprehensive assessment of various parameter configurations for the denoiser and selected the most suitable settings. Regarding the impact of $\mathbf{W}$, our results indicate that the choice of wavelet transform does not significantly affect the denoising performance. In contrast, we observed a significant enhancement in denoising performance by increasing the number of layers $L$ from $1$ to $10$. However, increasing $L$ beyond $10$ yields marginal improvements in the generalization capability.  Thus, we capped $L$ at $10$ to keep the model compact without compromising the denoising performance. Regarding stride $s$, our findings align with common trends in patch aggregation algorithms, namely, a higher stride is associated with decreased accuracy but reduced computational demands. We found that $s=8$ provides an optimal balance between accuracy and efficiency. Finally, the patch size plays a crucial role in the denoising performance. Keeping the stride constant, enlarging the patch size from $64$ to $128$ results in an accuracy improvement ranging from $0.10$ to $0.25$ dB and an enhancement in the SSIM score by $0.01$ to $0.02$. However, this adjustment also leads to an increase in the number of parameters of the denoiser.

\subsection{Deblurring}
\label{subsec:deblur}

The forward model for deblurring is $\y = \F\bxi + \boldeta$, where $\y$ is the blurred image, $\F$ is a blur operator, and $\boldeta$ is white Gaussian noise \cite{dong2011image}. For deblurring, the loss function is $f(\x)=\|\y -\F\x\|^2$, where $\F$ is the blur matrix. The gradient $\nabla f$ is $\beta$-Lipschitz. Specifically, we use a normalized blur for all experiments; as a result, we have $\beta=1$ ($\beta$ is the spectral norm of $\F$). For the deblurring experiments in Table \ref{counterexample1} and Fig.~\ref{propplots}, we took $\F$ to be a Gaussian blur of width $25 \times 25$ and standard deviation $1.6$ and the noise level is $\sigma=0.04$. For deblurring using PnP-FBS, the step size is set to $\gamma=0.001$ for RED-PG; for PnP-FBS, we used $\gamma=0.5$ which satisfies the condition $\gamma \leqslant 2/\beta$ in Corollaries~\ref{corr:Dcontractive} and \ref{corr:Daveraged}. For PnP-FBS, we used a noise level $\sigma=5$ for both our denoisers, and $\sigma=10$ for RED-PG.

In Table~\ref{counterexample1}, we present numerical evidence to show that, in the absence of a theoretical guarantee, simply plugging a denoiser (in this case DnCNN) into PnP-FBS can cause the iterations to diverge. Indeed, the distance between successive iterates $\lVert \x_k - \x_{k-1} \rVert$ is shown to diverge for DnCNN, which means that the sequence $(\x_k)$ does not converge. This results in spurious reconstructions, as evident from the PSNR values in Table \ref{counterexample1}. 
On the other hand, we notice that $\lVert \x_k - \x_{k-1} \rVert$ is decreasing for the proposed denoisers, thanks to their nonexpansive property. 

\begin{figure*}[!htp]
\captionsetup[subfigure]{labelformat=empty}

\centering
\subfloat[\hspace{-1.5cm}Input]{\rule{2cm}{0pt}}\quad
\subfloat[\hspace{-1cm}BM3D]{\rule{2cm}{0pt}}\quad
\subfloat[\hspace{-0.3cm}DnCNN]{\rule{2cm}{0pt}}\quad
\subfloat[\hspace{0.1cm}N-CNN]{\rule{2.2cm}{0pt}}\quad
\subfloat[\hspace{0.4cm}$\mathrm{D}_{\FBSop}$]{\rule{2cm}{0pt}}\quad
\subfloat[\hspace{1cm}$\mathrm{D}_{\DRSop}$]{\rule{2cm}{0pt}}
\vspace{-0.3cm}

\subfloat[ ($14.12$, $0.2940$)]{\includegraphics[width=0.16\linewidth]{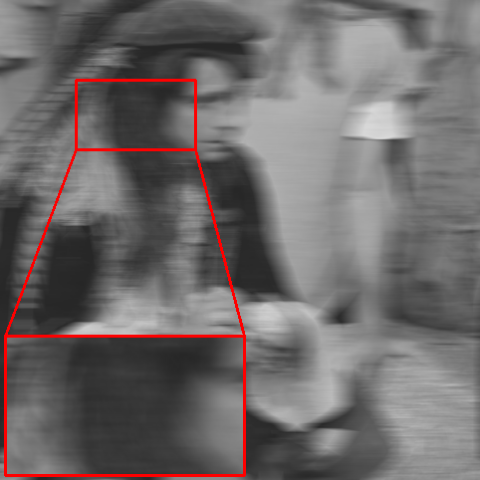}} \hspace{0.05mm}
\subfloat[ ($36.52$, $0.9458$)]{\includegraphics[width=0.16\linewidth]{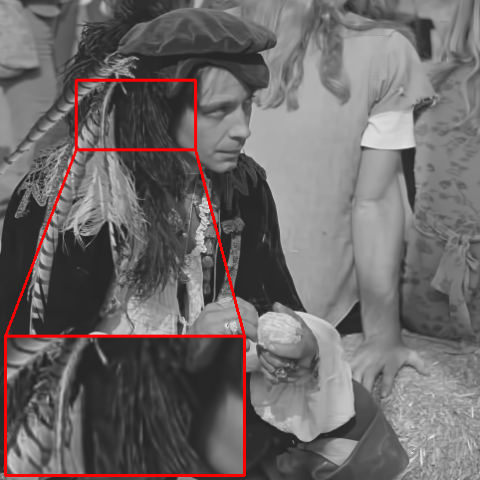}}  \hspace{0.05mm}
\subfloat[($36.21$, $0.9316$)]{\includegraphics[width=0.16\linewidth]{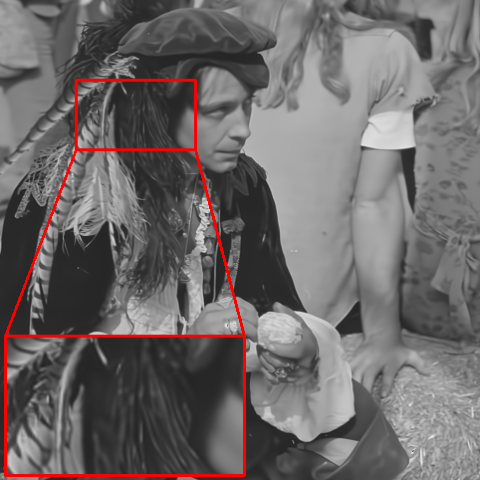}}  \hspace{0.05mm}
\subfloat[($32.73$,  $0.8885$)]{\includegraphics[width=0.16\linewidth]{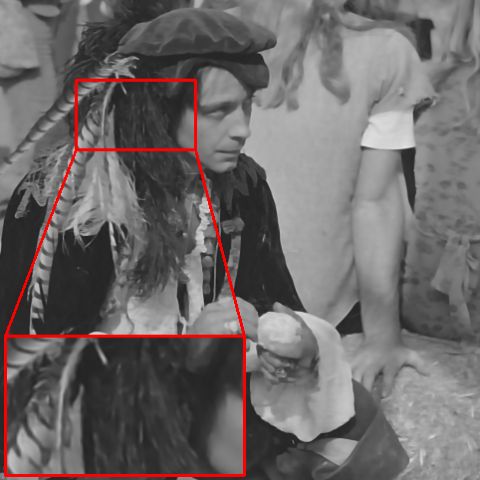}}  \hspace{0.05mm}
\subfloat[($\mathbf{36.79}$, $\mathbf{0.9511}$)]{\includegraphics[width=0.16\linewidth]{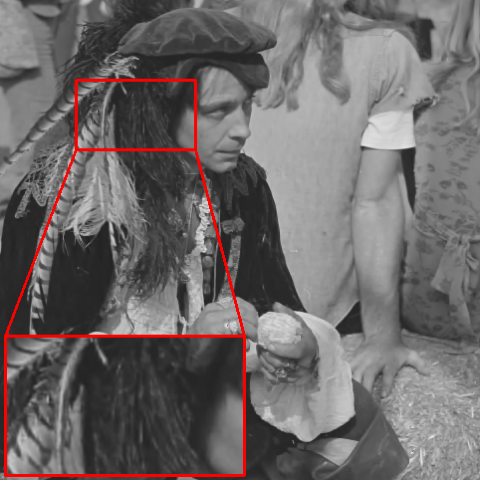}} \hspace{0.05mm}
\subfloat[($36.43$, $0.9465$)]{\includegraphics[width=0.16\linewidth]{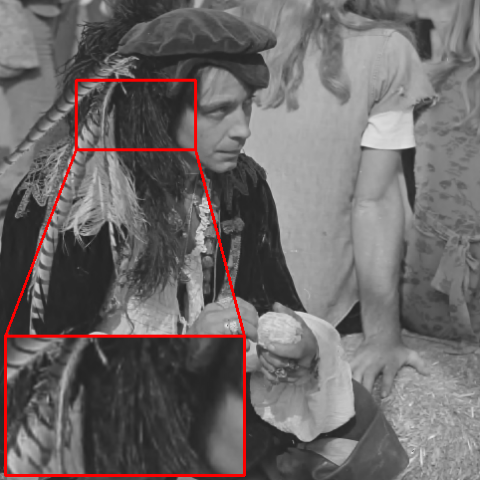}}  

\subfloat[($18.51$, $0.4142$)]{\includegraphics[width=0.16\linewidth]{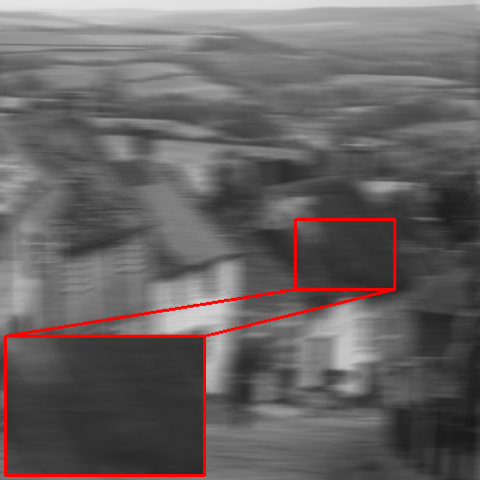}} \hspace{0.05mm}
\subfloat[(${36.23}$, $0.9287$)]{\includegraphics[width=0.16\linewidth]{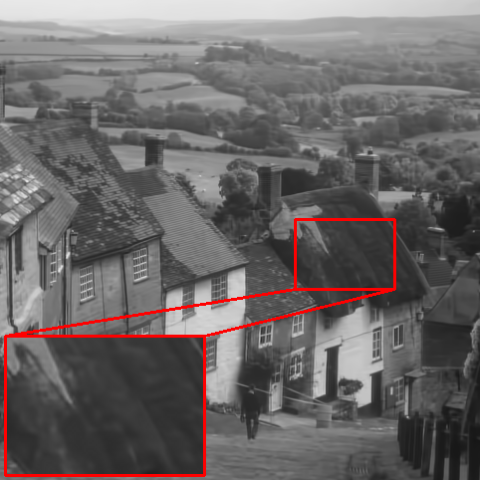}}  \hspace{0.05mm}
\subfloat[($35.27$, $0.9143$)]{\includegraphics[width=0.16\linewidth]{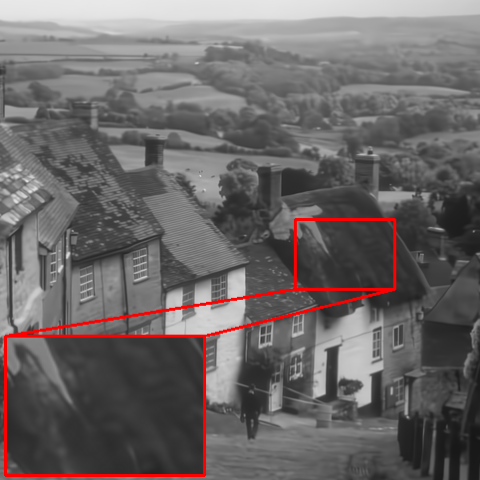}}  \hspace{0.05mm}
\subfloat[($32.91$, $0.8755$)]{\includegraphics[width=0.16\linewidth]{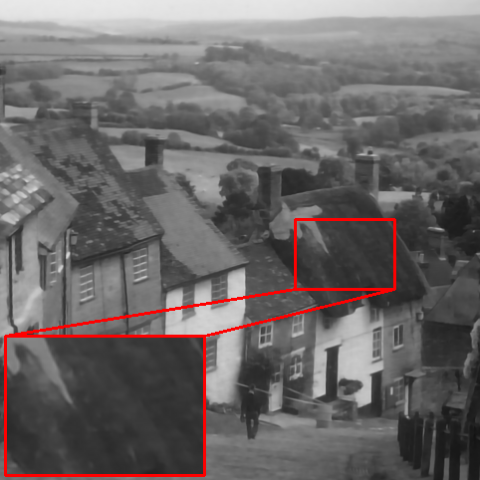}}  \hspace{0.05mm}
\subfloat[($\mathbf{36.50}$, $\mathbf{0.9422}$)]{\includegraphics[width=0.16\linewidth]{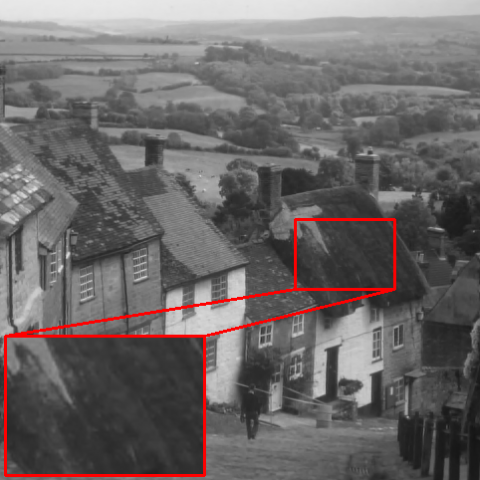}} \hspace{0.05mm}
\subfloat[($36.13$, $0.9365$)]{\includegraphics[width=0.16\linewidth]{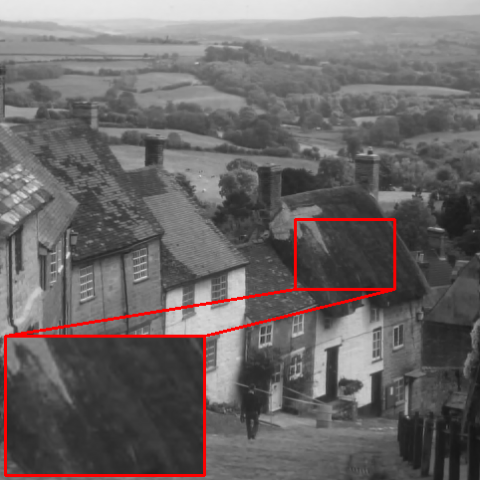}} 

\vspace{0.1cm}
\caption{\textbf{Deblurring using RED-PG} with different denoisers. The PSNR and SSIM values are displayed in the order (PSNR(dB), SSIM). The performance of our denoisers is comparable to BM3D and DnCNN. In particular, notice the fine details in the zoomed region where we get better detail enhancement from our denoisers.}
\label{debluringresults}
\end{figure*}

\begin{table*}[!htp]
\caption{\textbf{Deblurring using RED-PG} (PSRN/SSIM) for the blur \textbf{kernel 1} in \cite{levin2009understanding} (see text for more details). 
}
\centering
\renewcommand{\arraystretch}{1.2}
%\resizebox{2.0\columnwidth}{!}{ 
\begin{tabular}{  c|c c c c } 
\hline
&  DnCNN &  N-CNN &  $\D_{\mathrm{FBS}}$ &   $\D_{\mathrm{DRS}}$ \\
\hline
 {barbara} &  $\mathbf{25.08}$/$0.7307$  &  $23.98$/$0.6892$ &  $\mathbf{25.08}$/$\mathbf{0.7410}$ &  $24.98$/$0.7371$ \\
%\hline
 {ship} &  $27.39$/$0.7502$  &  $26.82$/$0.7145$ &  $\mathbf{27.52}$/$\mathbf{0.7627}$ &  $27.47$/$0.7600$ \\
%\hline
 {cameraman} &  $24.92$/$0.7867$  &  $\mathbf{25.01}$/$0.7700$ &  $24.96$/$\mathbf{0.7808}$ &  $24.91$/$\mathbf{0.7808}$ \\
%\hline
 {couple} &  $26.95$/$0.7346$  &  $26.50$/$0.7046$ &  $\mathbf{27.15}$/$\mathbf{0.7541}$ &  $27.09$/$0.7508$ \\
%\hline
 {fingerprint} &  $25.86$/$0.8773$  &  $23.55$/$0.7817$ &  $\mathbf{25.99}$/$\mathbf{0.8843}$ &  $25.90$/$0.8809$ \\
%\hline
 {hill} &  $28.41$/$0.7301$  &  $28.19$/$0.7125$ &  $\mathbf{28.71}$/$\mathbf{0.7573}$ &  $28.64$/$0.7532$ \\
%\hline
 {house} &  $\mathbf{30.29}$/$\mathbf{0.8310}$  &  $29.85$/$0.8252$ &  $29.92$/$0.8269$ &  $29.90$/$0.8278$ \\
%\hline
 {lena} &  $\mathbf{30.78}$/$\mathbf{0.8547}$  &  $30.36$/$0.8421$ &  $30.73$/$0.8540$ &  $30.68$/$0.8545$ \\
%\hline
 {man} &  $28.20$/$0.7749$  &  $27.94$/$0.7550$ &  $\mathbf{28.45}$/$\mathbf{0.7928}$ &  $28.38$/$0.7902$ \\
%\hline
 {montage} &  $24.56$/$\mathbf{0.8881}$  &  $25.10$/$0.8786$ &  $\mathbf{24.78}$/$0.8653$ &  $24.70$/$0.8683$ \\
%\hline
 {peppers} &  $\mathbf{24.96}$/$0.8227$  &  $24.95$/$0.8174$ &  $24.95$/$\mathbf{0.8255}$ &  $24.90$/$0.8252$ \\
\hline
\end{tabular}%}
\label{Debluringtab}
\end{table*}

Next, we present numerical evidence on PnP-FBS convergence using the proposed denoisers. This is done using a deblurring experiment on the Set 12 dataset (the settings are the same as in Table~\ref{counterexample1}). The results are shown in Fig.~\ref{propplots}~(a-b), where we plot $\lVert \x_k - \x_{k-1} \rVert$ and PSNR for different $k$. As expected, $\lVert \x_k - \x_{k-1} \rVert$ is decreasing and the PSNR stabilizes in less than $50$ iterations.  We also demonstrate that due to the contractivity of the FBS operator $\T_{\mathrm{PnP}}$ using our denoiser $\D_{\FBSop}$, the iterates converge to a unique fixed point even if we start with diverse initializations; the final reconstructions are identical, as shown in Fig~\ref{propplots}(c), where the MSE between the different reconstructions is of the order $1\mathrm{e}\mbox{-}8$.

We next compare the regularization capacity of our denoisers with BM3D, DnCNN, and N-CNN for deblurring, where $\F$ is a horizontal motion blur. This is done for the RED-PG algorithm. The results are shown for a couple of images in Fig.~\ref{debluringresults}. For the two images from Fig.~\ref{Inputimages}, which have a mix of delicate textures, smooth regions, and broader features (top row), the reconstruction from our denoisers exhibits finer details than other denoisers. 

A detailed comparison for RED-PG is provided in Table \ref{Debluringtab}, where $\F$ is one of the blur kernels from \cite{levin2009understanding}. We notice that our performance is comparable with DnCNN.

\begin{table}[!htp]
\caption{\textbf{Deblurring using PnP-FBS}: Comparison with the denoiser in \cite{repetti2022dual}. See Table~1 in \cite{repetti2022dual} for the notations A-E.}
\centering
\renewcommand{\arraystretch}{1.2}
%\resizebox{1.6\columnwidth}{!}{
\begin{tabular}{ c| c | c c c c c} 
\hline
{$\sigma$} & & A & B & C & D & E \\
\hline
\multirow{3}{*}{0.03} & Noisy & $12.78$ & $19.29$ & $16.06$ & $16.24$ & $22.54$ \\
& DualFB & $16.33$ & $24.81$ & $22.20$ & $20.41$ & $28.34$ \\
& $\D_{\DRSop}$ & $15.83$ & $23.97$ & $21.35$ & $20.28$ & $28.28$ \\
\hline
\multirow{3}{*}{0.05} & Noisy & $12.12$ & $18.11$ & $15.12$ & $14.58$ & $20.60$ \\
& DualFB & $14.97$ & $23.54$ & $20.65$ & $19.31$ & $27.47$ \\
& $\D_{\DRSop}$ & $14.42$ & $21.85$ & $18.42$ & $18.55$ & $25.60$\\
\hline
\multirow{3}{*}{0.08} & Noisy & $10.82$ & $16.11$ & $13.44$ & $12.10$ & $17.85$ \\
& DualFB & $14.49$ & $22.75$ & $19.68$ & $18.71$ & $26.93$ \\
& $\D_{\DRSop}$ & $13.74$ & $21.00$ & $17.83$ &  $17.89$ & $24.97$\\
%\hline
\hline
\end{tabular}%}
\label{DualFBcomparetab}
\end{table}

\begin{table}[!htp]
\caption{\textbf{Deblurring using PnP-FBS}: Comparison with the denoiser in \cite{hurault2022proximal}, averaged over CBSD68 dataset for $10$ different kernels} 
\centering
%\renewcommand{\arraystretch}{1.2}
%\resizebox{1.6\columnwidth}{!}{
\begin{tabular}{ c | c c c } 
\hline
\multirow{2}{*}{Method}& \multicolumn{3}{c}{$\sigma$}\\
 & $0.01$ & $0.03$ &  $0.05$\\
\hline 
Prox-PnP  & $30.57$ & $27.80$ & $26.61$ \\
\hline 
Proposed & $28.50$ & $27.11$ & $26.43$\\
\hline
\end{tabular}%}
\label{ProxPnPPGDcomparetab}
\end{table}

In Table \ref{DualFBcomparetab}, we compare $\D_{\DRSop}$ with the denoiser in \cite{repetti2022dual}, which we term as ``DualFB''. 
%The denoiser in \cite{repetti2022dual} is obtained by applying FBS to a denoising problem with box constraints on the reconstruction. The difference with our approach is that FBS is applied to the Fenchel dual of the denoising problem. Moreover, unlike our proposal where wavelet transform is fixed, the weights of linear transform are optimized during training in \cite{repetti2022dual}.
We replicated the experimental setup used for the deblurring experiment in \cite[Table~1]{repetti2022dual} --- an asymmetric blur operation followed by the addition of noise at various levels. For a fair comparison, similar to the experiment in \cite{repetti2022dual}, we use PnP-FBS as the reconstruction algorithm for our denoiser and signal-to-noise ratio (SNR) as the performance metric.  We see from  Table~\ref{DualFBcomparetab} that our denoiser can closely parallel the performance of DualFB. In fact, it is not surprising that DualFB yields superior results since the linear layers are optimized in \cite{repetti2022dual}, whereas we use fixed handcrafted transforms. 

We next compare $\D_{\DRSop}$ with the  ``Prox-PnP'' denoiser from \cite{hurault2022proximal}; we perform the comparison by using them as regularizers within the PnP-FBS framework. We replicated the experimental setup used for the deblurring experiment in \cite[Table~3]{hurault2022proximal} in which the PSNR is averaged over the CBSD68  dataset for $10$ different kernels and different noise levels. As shown in Table~\ref{ProxPnPPGDcomparetab}, our denoiser exhibits comparable reconstruction capacity to Prox-PnP, though our gray-scale denoiser is applied channel-by-channel. It is worth noting that our denoiser is contractive by construction, whereas the denoiser in  \cite{hurault2022proximal} is ``softly'' enforced to be $1$-Lipschitz using a loss-based penalization. We assume that training our denoiser specifically for color images can boost the color deblurring performance.

In light of the above comparisons, we wish to emphasize that the key advantage of our denoiser is that it comes with convergence guarantees. On the other hand, the convergence result in \cite[Proposition~2]{repetti2022dual} assumes that the same linear transform is used across all layers, and this is generally not true if we optimize the transforms. It remains to be seen if it is possible to optimize our construction to match the denoiser in \cite{repetti2022dual}.

\begin{figure*}[!htp]
\centering
\subfloat[\footnotesize BM3D.]{\includegraphics[width=0.33\linewidth]{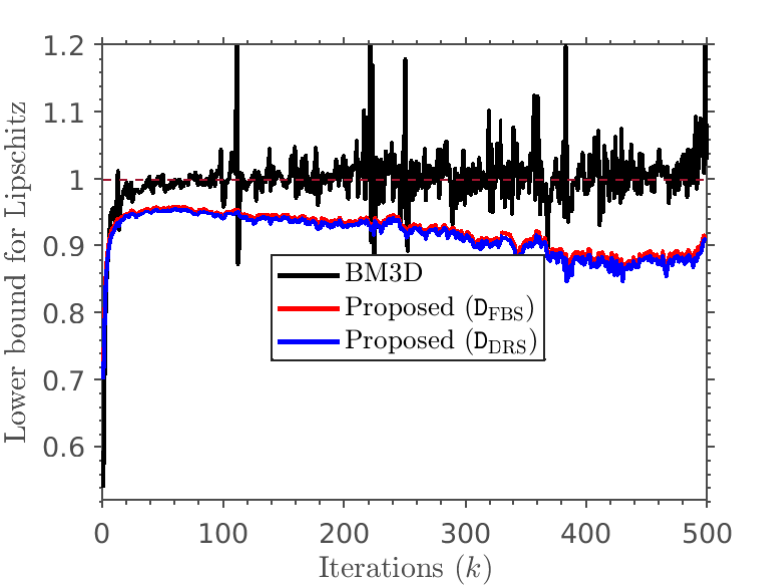}} \hspace{0.05mm}
\subfloat[\footnotesize DnCNN.]{\includegraphics[width=0.33\linewidth,height=0.252\linewidth]{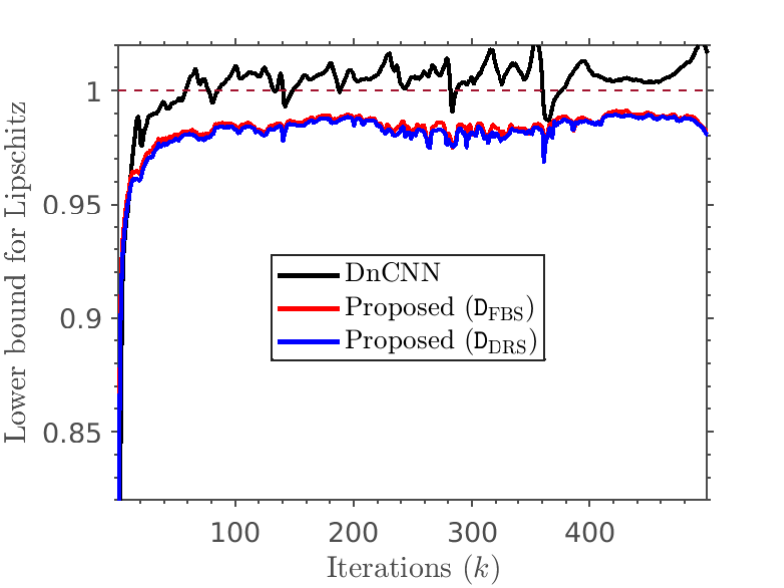}} \hspace{0.05mm}
% \subfloat[\footnotesize SimpleCNN.]{\includegraphics[width=0.23\linewidth]{nonexpansiveplots/SimpleCNN_nonexpansive.pdf}} \hspace{0.05mm}
\subfloat[\footnotesize N-CNN \cite{Ryu2019_PnP_trained_conv}]{\includegraphics[width=0.33\linewidth,height=0.235\linewidth]{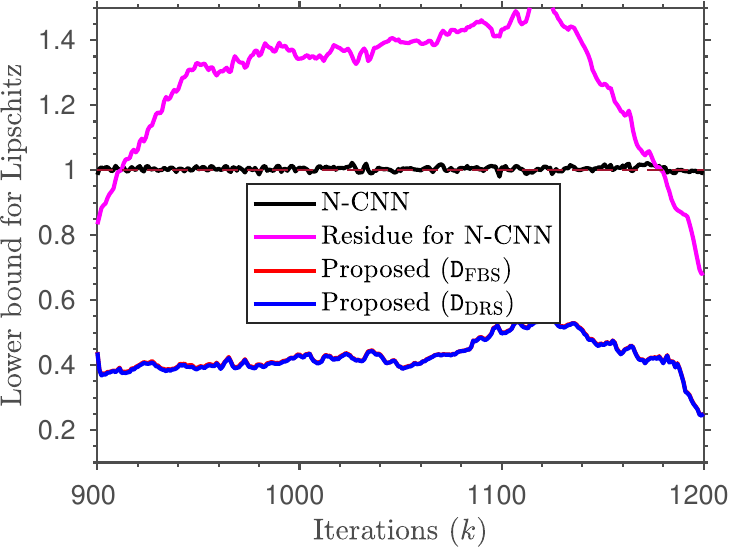}}
\caption{Violation of nonexpansive property for BM3D, DnCNN and N-CNN. We perform image superresolution with different denoisers plugged into PnP-FBS. A lower bound for the Lipschitz constant of the respective denoisers is computed from the data encountered in the PnP pipeline. More precisely, since the  Lipschitz constant is defined to be the supremum of $\|\D(\u) -\D(\v) \| / \| \u - \v \|$ over for all possible inputs $\u$ and $\v$, setting $\u=\x_k$ and $\v=\x_{k+1}$ we obtain a lower bound for the Lipschitz constant at different iterations. Note that for BM3D, DnCNN, and N-CNN, the lower bound is $>1$ in many iterations, proving that these denoisers are not nonexpansive. As expected, the lower bound for the Lipschitz constant of the proposed denoisers is within $1$.
}
\label{nonexpansiveplots}
\end{figure*}

\begin{figure*}[!htp]
\captionsetup[subfigure]{labelformat=empty}
\centering
\subfloat[Bicubic]{\includegraphics[width=0.22\linewidth]{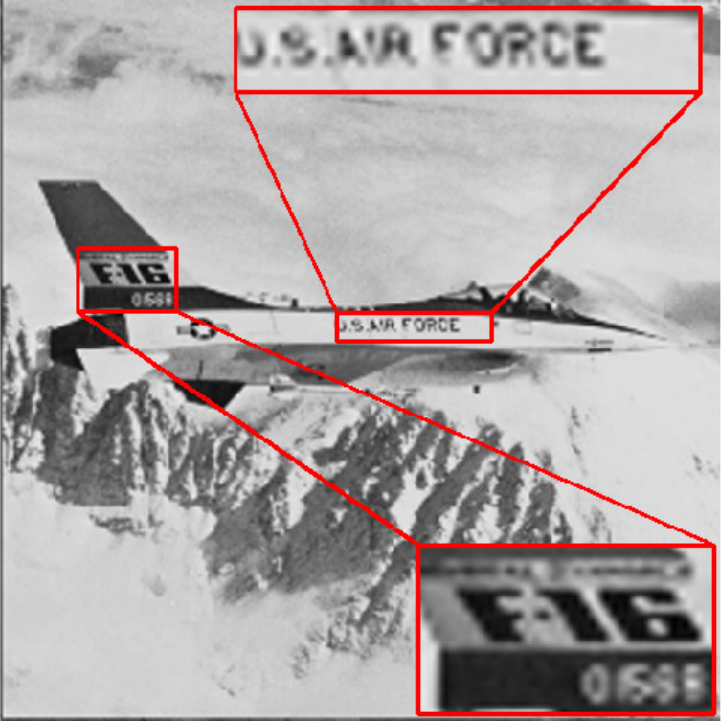}} \hspace{0.05mm}
\subfloat[\scriptsize BM3D ($\bf{31.03}$, $\bf{0.9250}$)]{\includegraphics[width=0.22\linewidth]{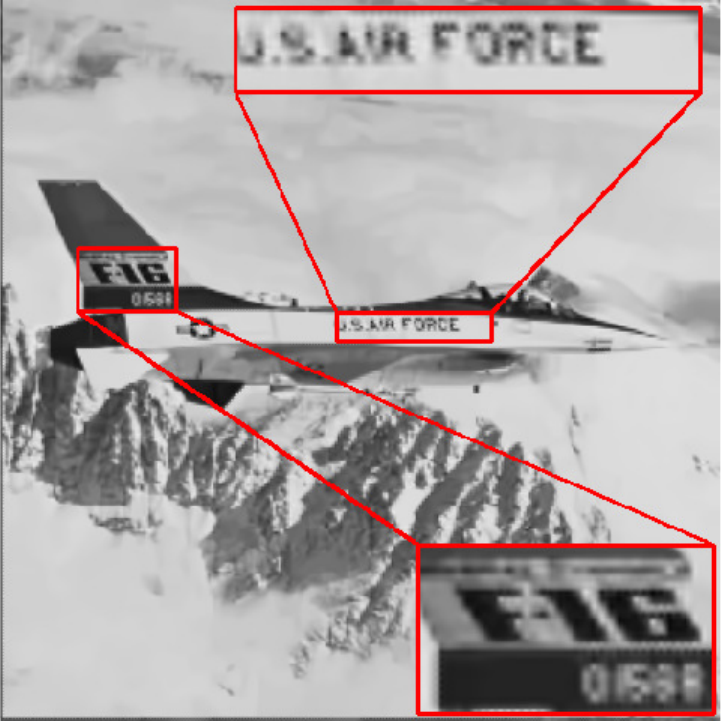}} \hspace{0.05mm}
\subfloat[\scriptsize DnCNN ($30.22$, $0.9167$)]{\includegraphics[width=0.22\linewidth]{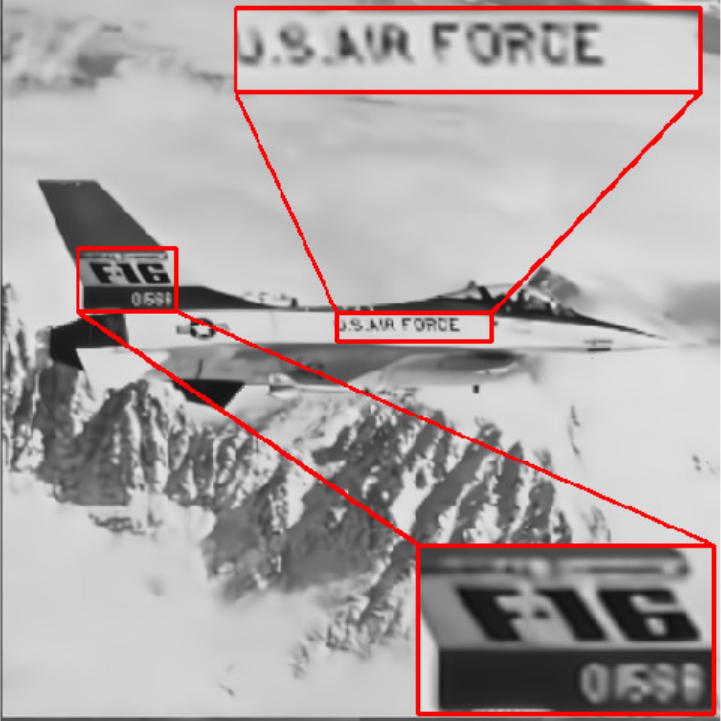}} \hspace{0.05mm}
\subfloat[\scriptsize N-CNN ($28.92$, $0.8925$)]{\includegraphics[width=0.22\linewidth]{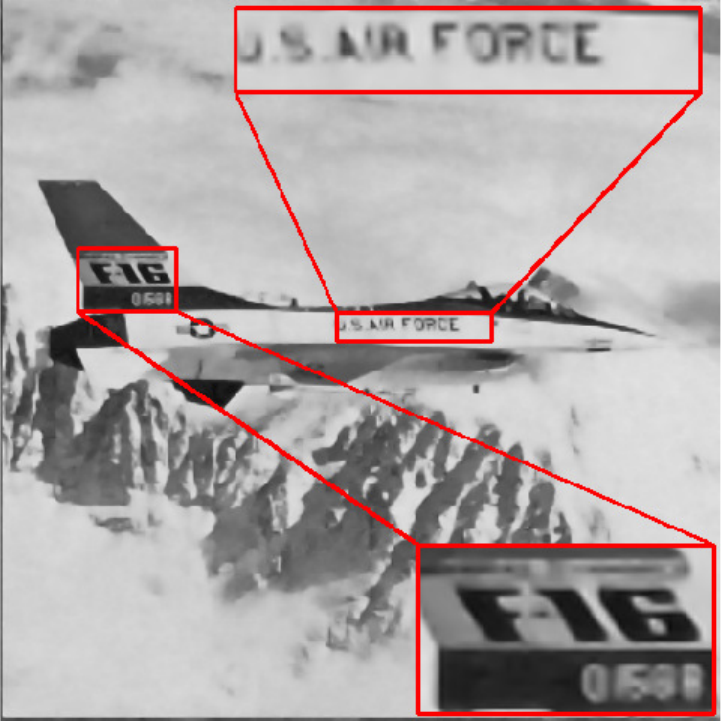}} 

\subfloat[\scriptsize $\mathrm{D}_{\FBSop}$ ($30.61$, $0.9167$)]{\includegraphics[width=0.22\linewidth]{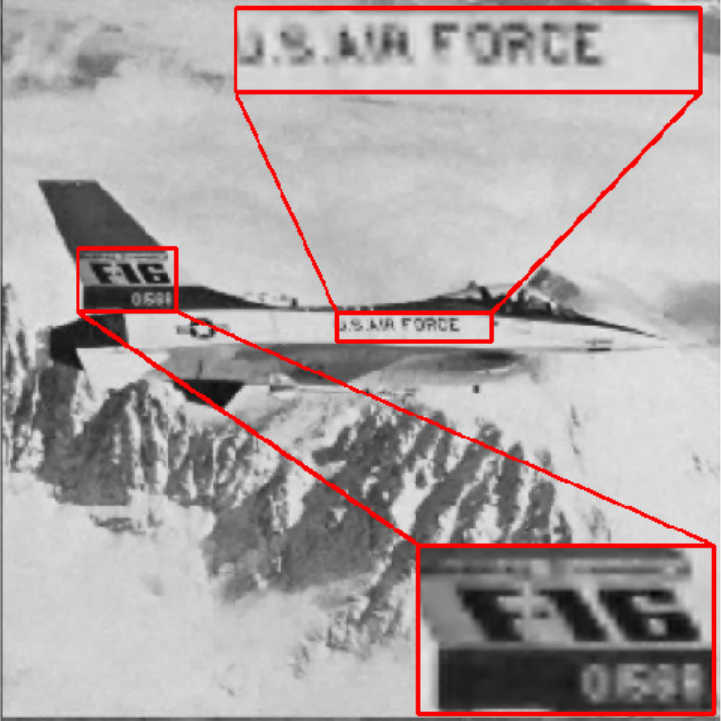}} \hspace{0.05mm}
\subfloat[\scriptsize $\mathrm{D}_{\DRSop}$ ($30.48$, $\bf{0.9202}$)]{\includegraphics[width=0.22\linewidth]{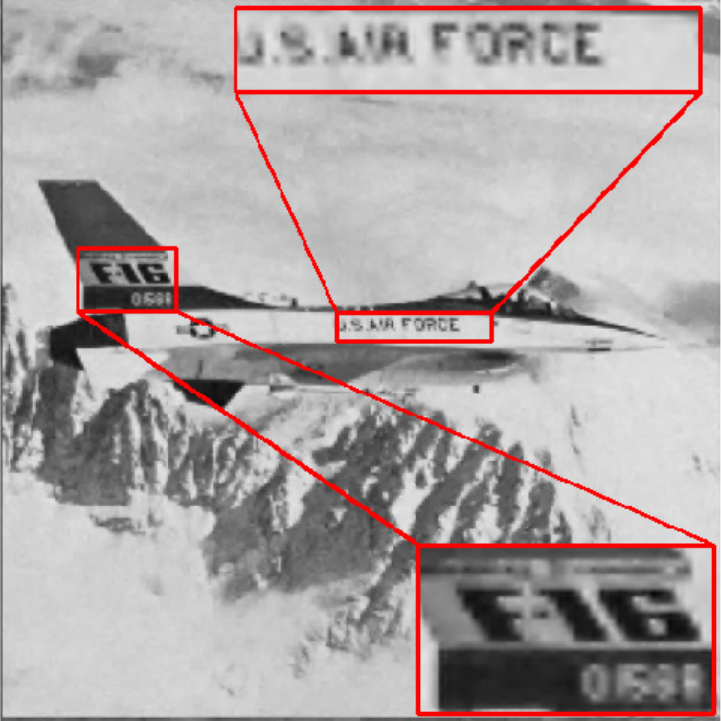}} \hspace{0.1mm}
\subfloat[ Convergence plot]{\includegraphics[width=0.3\linewidth]{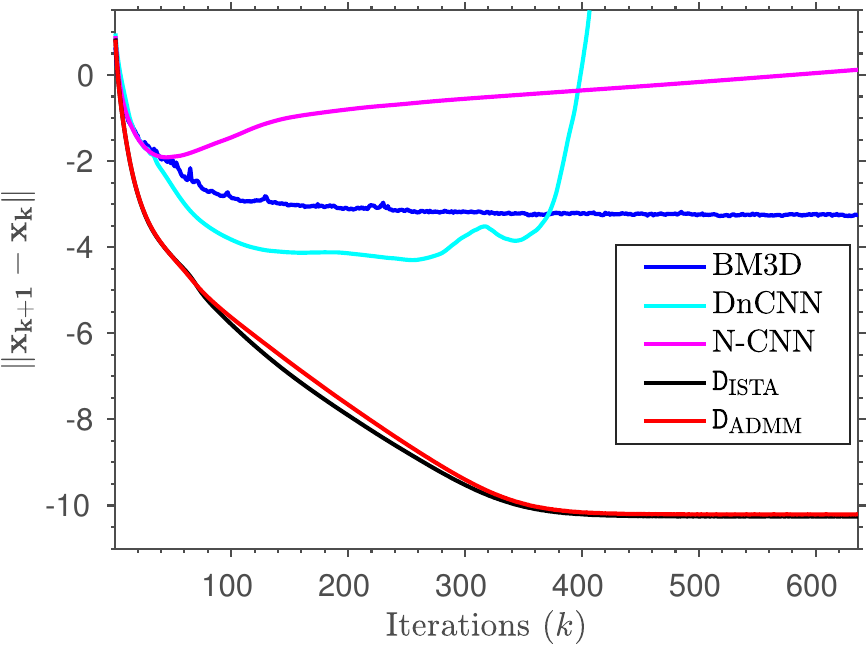}} 
\caption{\textbf{Superresolution using PnP-FBS} with different denoisers. The PSNR and SSIM after $30$ iterations are displayed. Reconstructions obtained from our denoisers ($\mathrm{D}_{\FBSop}$ and $\mathrm{D}_{\DRSop}$) are comparable with BM3D and DnCNN. The convergence plot suggests that the PnP iterates diverge for DnCNN and N-CNN.}
\label{superresolutionresults}
\end{figure*}

\subsection{Superresolution} 
\label{subsec:SR}

A widely used model for superresolution is $\y = \S \B \bxi + \boldeta,$ where $\B \in \Re^{n \times n}$ and $\S \in \Re^{m \times n}$  are the blur and subsampling operators ($m = n/k$, $k > 1$), and $\boldeta$ is Gaussian noise \cite{chan2017plug,nair2021fixed}. For superresolution, the loss function is $f(\x)=\|\y -\S\B\x\|^2$, where $\S$ is a decimation matrix and $\B$ is a blur matrix.  As in deblurring, we used a normalized blur. In this case, $\beta$ is bounded by the product of the spectral norms of $\S$ and $\B$, so we have $\beta\leqslant1$. In PnP-FBS,  we used the step size $\gamma=1.0$ that satisfies the condition $\gamma \leqslant 2/\beta$ in Corollaries~ \ref{corr:Dcontractive} and \ref{corr:Daveraged}. We set the noise level to $\sigma=5$ for both our denoisers for all experiments.

\begin{table*}[!htp]
\caption{Residual $\|\x_k - \x_{k-1}\|$ for PnP-FBS with different denoisers.}
\centering
\renewcommand{\arraystretch}{1.2}
%\resizebox{.45\textwidth}{!}{
\begin{tabular}{c|c c c c c c }
\hline
\multirow{2}{*}{denoiser} & \multicolumn{6}{c}{iterations ($k$)} \\ 
& $10$ & $50$ & $100$ & $500$ & $1000$ & $3000$\\
\hline
 N-CNN \cite{Ryu2019_PnP_trained_conv} & $0.54$ & $0.20$ & $0.19$ & $0.36$ & $23.45$ & $3\mathrm{e}$+$12$\\
 $\D_{\mathrm{FBS}}$ & $0.31$ & $0.15$ & $0.11$ & $0.0602$ & $0.0513$ & $2\mathrm{e}$-$10$\\
 $\D_{\mathrm{DRS}}$ & $0.32$ & $0.16$ & $0.11$ & $0.0658$ & $0.0562$ & $2\mathrm{e}$-$10$\\
\hline
\end{tabular}%}
\label{counterexample2} 
\end{table*}

\begin{table*}
\caption{\textbf{Superresolution using PnP-FBS}: Comparison of the reconstruction capacity of different denoisers. (see text for details)}
\centering
\renewcommand{\arraystretch}{1.2}
%\resizebox{1.0\columnwidth}{!}{ 
\begin{tabular}{ c|c c c c} 
\hline
&  DnCNN &  N-CNN &  $\D_{\mathrm{FBS}}$ &    $\D_{\mathrm{DRS}}$\\
\hline
 cameraman &  $\mathbf{27.12}$/$0.8519$  &  $26.89$/$0.8450$ &  $26.60$/${0.8616}$ &  $26.63$/$\mathbf{0.8540}$ \\
 house &  ${33.29}$/$0.8695$  &  $32.84$/$0.8736$ &  $32.75$/$0.8531$ &  $\mathbf{33.70}$/$\mathbf{0.8780}$ \\
 peppers &  $\mathbf{28.85}$/$0.9071$  &  $28.43$/$0.9027$ &  $27.74$/$\mathbf{0.9079}$ &  $27.67$/$0.9059$ \\
 {starfish} &  $29.87$/$0.9014$  &  $27.59$/$0.8609$ &  $28.80$/$0.9041$ &  $\mathbf{28.87}$/$\mathbf{0.8979}$ \\
 {butterfly} &  $\mathbf{30.28}$/$0.9403$  &  $30.36$/$0.9340$ &  $28.81$/$0.9341$ &  $28.95$/$\mathbf{0.9318}$ \\
 {plane} &  $\mathbf{26.77}$/$0.8734$  &  $25.97$/$0.8520$ &  $26.07$/$0.8770$ &  $26.23$/$\mathbf{0.8735}$ \\
 {parrot} &  $\mathbf{28.18}$/$0.8866$  &  $27.60$/$0.8865$ &  $27.62$/$0.8998$ &  $27.59$/$\mathbf{0.8934}$ \\
 {lena} &  $34.00$/$\mathbf{0.8980}$  &  $33.21$/$0.8942$ &  $33.87$/$0.9118$ &  $\mathbf{33.76}$/$\mathbf{0.9063}$ \\
%\hline
 {barbara} &  $24.87$/$0.7623$  &  $24.52$/$0.7360$ &  ${25.00}$/${0.7775}$ &  $\mathbf{24.90}$/$\mathbf{0.7629}$ \\
%\hline
 {ship} &  $30.29$/$0.8414$  &  $29.33$/$0.8065$ &  ${30.02}$/${0.8531}$ &  $\mathbf{29.93}$/$\mathbf{0.8430}$ \\
%\hline
 {man} &  $30.79$/$0.8595$  &  $30.25$/$0.8409$ &  ${30.83}$/${0.8800}$ &  $\mathbf{30.67}$/$\mathbf{0.8678}$ \\
%\hline
 {couple} &  $29.96$/$0.8436$  &  $28.98$/$0.8105$ &  ${29.57}$/${0.8522}$ &  $\mathbf{29.51}$/$\mathbf{0.8440}$ \\
\hline

%\hline

\hline
\end{tabular}%}
\label{Superrestab}
\end{table*}

For the results in Fig.~\ref{nonexpansiveplots} and Table \ref{counterexample2}, $\B$ is a Gaussian blur of size $25 \times 25$ and standard deviation $1.6$, and $\sigma=0.04$. We start by presenting numerical evidence which shows that BM3D, DnCNN, and N-CNN are not nonexpansive, even if we restrict the input to images encountered in the PnP pipeline. More precisely, we apply PnP-FBS for image superresolution and compute the output-input ratio $\| \x_{k+1} - \x_{k}\| / \| \z_{k+1} - \z_{k} \|$ for the denoising operation $\x_{k+1}=\D(\z_{k+1})$. This gives a lower bound for the Lipschitz constant of the plugged denoiser. In particular, if any of these ratios is $>1$, the Lipschitz constant cannot be less than $1$, and the denoiser cannot be nonexpansive. The ratio for different denoisers across iterations is shown in Fig.~\ref{nonexpansiveplots}. We notice that the ratio is $>1$ at many iterations for BM3D, DnCNN, and N-CNN, proving that the denoisers are not nonexpansive. For N-CNN, the residual is constrained to be nonexpansive, but this constraint is also violated, as shown in Fig.~\ref{nonexpansiveplots}(c). On the other hand, the ratio is always $<1$ for our denoiser, which is consistent with the theoretical results in Section \ref{proposed}. Similar to the counterexample in Table \ref{counterexample1}, we provide a counterexample in Table  \ref{counterexample2} that suggests divergence of the iterates on plugging N-CNN into PnP-FBS. 
%The parameter $\rho$ is fixed to be $0.001$, which is in line with convergence results of PnP-FBS . 
This is consistent with the observation in Fig.~\ref{nonexpansiveplots} that N-CNN violates the nonexpansivity property.

We next compare our denoisers with BM3D, DnCNN, and N-CNN for the PnP-FBS algorithm using just downsampling ($\B$ is the identity operator). The results are shown in Fig.~\ref{superresolutionresults} at $\sigma=5/255$.  
%The number of iterations used to output the restoration estimates is $30$. 
It is clear from the visual results and the metrics that the restoration results using our denoisers are comparable with BM3D and DnCNN. We also report the residual $\lVert \x_k - \x_{k-1} \rVert$ for sufficiently large $k$. This monotonically goes to zero for our denoisers but is seen to diverge for DnCNN and N-CNN. A detailed comparison for the PnP-FBS is shown in Table \ref{Superrestab}, where the blur and noise level are the same as in Fig.~\ref{nonexpansiveplots}. Again, we notice that our performance is comparable with DnCNN.

In summary, the findings from our empirical study are as follows:
\begin{enumerate}
\item The numerical results in Fig.~\ref{nonexpansiveplots}, and  Tables~\ref{counterexample1} and \ref{counterexample2}, show that denoisers such as BM3D, DnCNN, and N-CNN are not guaranteed to be nonexpansive and can cause PnP algorithms to breakdown.
\item Fig ~\ref{propplots} demonstrates that the proposed denoisers exhibit empirical convergence in agreement with the theoretical predictions in Corollaries \ref{corr:Dcontractive} and  \ref{corr:Daveraged}.
\item From the results in Figures~\ref{debluringresults} and \ref{superresolutionresults}, and the extensive comparisons in Tables \ref{Denoisingtab}, \ref{Debluringtab}, and \ref{Superrestab}, we can conclude that the regularization capacity of the proposed denoisers is comparable with BM3D and DnCNN for deblurring and superresolution.
\end{enumerate}

\section{Conclusion}

We trained averaged and contractive patch denoisers by unfolding FBS and DRS applied to wavelet denoising. We extended them for image denoising using patch aggregation while preserving their averaged (contractivity) properties. Moreover, by using sufficiently many layers, we showed that their regularization capacity for PnP and RED can be brought to par with BM3D and DnCNN. Unlike existing CNN denoisers, we could guarantee convergence of PnP and RED using the proposed denoisers. We also presented numerical evidence to back these results. To our knowledge, this is the first work to exploit the properties of proximal operators in unfolded algorithms to develop %averaged and 
contractive denoisers. In future work, we wish to go beyond classical wavelet denoising and develop unfolded denoisers with superior denoising and regularization capacity. 

\section{Funding}

K.~N.~Chaudhury was supported by research grants CRG/2020/000527 and STR/2021/000011 from the Science and Engineering Research Board, Government of India.
%\begin{appendices}

\section{Appendix} 
\label{proof:imagedenoiser}

We prove Proposition~\ref{finalprop} in this section. First, we recall some notations from Section \ref{sec:imgdenoiser}.
\begin{enumerate}
\item For integers $p \leqslant q$, we denote $[p,q]=\{p,p+1,\ldots,q\}, \, [p,q]_s=\{ps,(p+1)s,\ldots, qs\}, \, [p,q]^2=[p,q] \times [p,q]$, and $[p,q]_s^2=[p,q]_s \times [p,q]_s$.

\item The input to the denoiser is an image $\X: \Omega \to \Re$, where $\Omega=[0,q-1]^2$. We also consider the image in matrix form as $\X \in \Re^{q \times q}$. 

\item We use circular shifts: for $\i \in \Omega$ and $\btau  \in \mathbb{Z}^2$, 
$\i - \btau$ is defined as $\left( (i_1 - \tau_1) \ \mathrm{mod} \ q, (i_2 - \tau_2) \ \mathrm{mod} \ q\right)$.

\item  We extract patches with stride $s$. More precisely, the starting coordinates of the patches are $$\J = [0, (q_s-1)]_s^2,$$
where $q_s =  q/s $. The cardinality of $\J$ is $q_s^2$ (total number of patches), and each pixel belongs to $k_s^2$ patches, where $k_s = k/s$.   

\item For $\i \in \Omega$, the patch operator $\Pcal_{\i}: \Re^{q \times q} \to \Re^{p}$ is defined in \eqref{patchop}.

\end{enumerate}

We now elaborate on the definition of the adjoint in \eqref{imagedenoiser}. For a patch vector $\z \in \Re^p$, if we let $\Y=\Pcal^*_{\i}(\z)$, then $\Pcal_{\i}(\Y) = \z$ and $\Y(\j) = 0$ for all $\j \notin \Omega_{\i}$, where $$\Omega_{\i}= \left\{\i + \btau: \ \btau \in [0,k-1]^2\right\}.$$ 
This completely specifies $\Y$. In particular, we have the following property of an adjoint: for all $\X \in \Re^{q \times q}$ and $\z \in \Re^{p}$,
\begin{equation}
\label{def:adjoint}
\langle \z, \Pcal_{\i}(\X) \rangle_{\Re^p} = \langle \Pcal_{\i}^*(\z), \X \rangle_{\Re^{q \times q}},
\end{equation}
where the inner product on the left (resp.~right) is the standard Euclidean inner product on $\Re^p$ (resp.~$\Re^{q \times q}$).

The main idea behind Proposition~\ref{finalprop} is to express the original system of overlapping patches in terms of non-overlapping patches. In particular, let $q_k=q/k$; since we assume $q$ to be a multiple of $k$, $q_k$ is an integer. Let
\begin{align*}
\J_0=[0, (q_k-1)]_k^2.
\end{align*}

By construction, $\J_0 \subseteq \J$ and the points in $\J_0$ are the starting coordinates of non-overlapping patches.  
It is not difficult to see that $\J$ is the disjoint union of  (circular) shifts of $\J_0$:
\begin{equation}
\label{disjointsets}
\J =  \bigcup_{\i \in [0,k_s-1]^2} \, \big( \J_0 + s \i  \big),
\end{equation}
where $[0,k_s-1]:=\{0, 1, \ldots, k_s-1\}$.

Using \eqref{disjointsets}, we can decompose \eqref{imagedenoiser} as follows:
\begin{equation}
\label{avg}
\D  = \frac{1}{k_s^2} \sum_{ \i \in [0, k_s-1]^2 } \D_{\i}, 
\end{equation}
where 
\begin{equation}
\label{defDi}
\D_{\i} = \sum_{\j \in \J_0} \ \left(\Pcal^*_{\j+s\i} \circ \D_{\mathrm{patch}} \circ \Pcal_{\j+s\i}\right).
\end{equation}
%Note that we assume that $s$ divides $k$ and $k$ divides $q$.

The proof of Proposition~\ref{finalprop} follows from a couple of observations. The first of these is an observation about operator averaging. We will say that operator $\T$ is $L$-contractive if its Lipschitz constant is at most $L$.

\vspace{-3mm}
\begin{lemma}
\label{avgavgfinite}
Let $\T_1, \ldots, \T_k$ be $L$-contractive. Then, their average,
\begin{equation}
\label{avgop}
\frac{1}{k} (\T_1 + \cdots + \T_k),
\end{equation}
 is $L$-contractive. On the other hand, if $\T_1,  \ldots, \T_k$ are $\theta$-averaged, then \eqref{avgop} is a  $\theta$-averaged operator.
\end{lemma}
\vspace{-3mm}

The second observation relates the contractivity (averaged) property of the patch denoisers to the image denoiser \eqref{defDi}.
\vspace{-3mm}
\begin{lemma}
\label{lemma:patchtoimg}
For $\i \in [0,k_s-1]^2$, the following hold.
\begin{enumerate}
\item If $\D_{\mathrm{patch}}$ is $L$-contractive, then $\D_{\i}$ is $L$-contractive.
\item If $\D_{\mathrm{patch}}$ is $\theta$-averaged, then $\D_{\i}$ is $\theta$-averaged.
\end{enumerate}
\end{lemma}
\vspace{-3mm}

To establish Proposition~\ref{finalprop}, note from \eqref{avg} that $\D$ is the average of $\{\D_{\i}\}$. The desired conclusion follows immediately from Lemma \ref{avgavgfinite} and \ref{lemma:patchtoimg}. 

We now give the proofs of Lemma \ref{avgavgfinite} and \ref{lemma:patchtoimg}. 

For Lemma \ref{avgavgfinite}, the second part follows from \cite[Proposition~4.30]{bauschke2017convex}; however, we give the proofs of both parts for completeness.

\begin{proof}[Proof of Lemma \ref{avgavgfinite}]
Let $\T=(1/k)(\T_1 + \cdots + \T_k)$, where $\T_1, \ldots, \T_k$ are $L$-contractive. Then, for $i=1,\ldots,k$ and for all $\x,\x' \in \R^n$,
\begin{equation}
\label{eq1}
\lVert \T_i(\x)  - \T_i(\x') \rVert \leqslant L \, \lVert \x - \x' \rVert. 
\end{equation}
 We can write
\begin{equation*}
\T(\x)  - \T(\x') = \frac{1}{k} \sum_{i=1}^k \, \left(\T_{i}(\x) - \T_{i}(\x') \right).
\end{equation*}
Using triangle inequality and \eqref{eq1}, we get
\begin{equation*}
\lVert \T(\x)  - \T(\x') \rVert  \leqslant  \frac{1}{k} \sum_{i=1}^k \, L \,  \lVert \x - \x' \rVert =  L \lVert \x - \x' \rVert.
\end{equation*}
This establishes the first part of Proposition~\ref{avgavgfinite}.

Next, suppose that $\T_1,  \ldots, \T_k$ are $\theta$-averaged, i.e., there exists nonexpansive operators $\N_1,\ldots,\N_k$ such that 
\begin{equation*}
\T_i= (1-\theta) \I +\theta \N_i \qquad (i=1,\ldots,k).
\end{equation*}
Then 
\begin{equation*}
\T = \frac{1}{k}(\T_1 + \cdots + \T_k)=  (1-\theta) \I  + \theta \N,
\end{equation*}
where $\N=(1/k)(\N_1 + \cdots + \N_k)$. 

We can again use triangle inequality to show that $\N$ is nonexpansive. Thus, we have shown that $\T$ is $\theta$-averaged, which completes the proof.
\end{proof}

We need the following result to establish Lemma \ref{lemma:patchtoimg}.

\vspace{-3mm}
\begin{lemma}
\label{lemma:ortho}
For any fixed $s$ and $\i \in [0, k_s-1]^2$, we have the following orthogonality properties:
\begin{enumerate}
\item For all $\X: \Omega \to \Re$, 
\begin{equation}
\label{equality2}
\sum_{\j \in \J_0} \ \lVert \Pcal_{\j+s\i}(\X) \rVert^2 = {\lVert \X \rVert}^2.
\end{equation}
We can equivalently write this as
\begin{equation}
\label{Xidentity}
\sum_{\j \in \J_0} \ \Pcal_{\j+s\i}^* \Pcal_{\j+s \i}=\I,
\end{equation}
where $\I$ is the identity operator on $\R^{q\times q}$. 
\item For all $\{\z_{\j} \in \R^p: \, \j \in \J_0\}$, 
\begin{equation}
\label{equality1}
 \big\lVert \sum_{\j \in \J_0} \ \Pcal_{\j+s \i}^*(\z_{\j}) \big\rVert^2 = \sum_{\j \in \J_0}\ \lVert \z_{\j} \rVert^2.
\end{equation}
\end{enumerate}
\end{lemma}
\vspace{-3mm}

It is understood that the norm on the left in \eqref{equality2} is the Euclidean norm on $\Re^{p}$, while the norm on the right is the Euclidean norm on $\Re^{q \times q}$.

We can establish Lemma \ref{lemma:ortho} using property \eqref{def:adjoint} and the fact that $\J_0$ are the starting coordinates of non-overlapping patches. We will skip this straightforward calculation. We are ready to prove Lemma \ref{lemma:patchtoimg}.

\begin{proof}[Proof of Lemma \ref{lemma:patchtoimg}]
Let $\D_{\mathrm{patch}}$ be $L$-contractive, i.e., for all $\z,\z' \in \R^p$,
\begin{equation}
\label{Dpatchcontract}
\lVert \D_{\mathrm{patch}}(\z) - \D_{\mathrm{patch}}(\z') \rVert \leqslant L \lVert \z-\z' \rVert.
\end{equation}
We have to show for all $\X, \X' \in \R^{q \times q}$,
\begin{equation}
\label{bound}
\lVert \D_{\i}(\X) - \D_{\i}(\X') \rVert^2  \leqslant L^2 \lVert \X-\X' \big\rVert^2.
\end{equation}
From \eqref{equality2} and \eqref{equality1}, we can write
\begin{align}
\label{mainproofineq}
\lVert \D_{\i}(\X) - \D_{\i}(\X') \rVert^2  = \sum_{\j \in \J_0}{\| \D_{\mathrm{patch}} (\z_{j}) -\D_{\mathrm{patch}} (\z'_{j}) \|}^2 \nonumber, 
\end{align}
where $\z_{j} = \Pcal_{\j+s\i}(\X)$ and $\z'_{j} = \Pcal_{\j+s\i}(\X')$. From \eqref{Dpatchcontract},
\begin{align*}
\lVert  \D_{\mathrm{patch}} (\z_{\j}) -\D_{\mathrm{patch}} (\z'_{\j}) \rVert 
&\leqslant L \, \lVert \z_{\j} -\z'_{\j}  \rVert \\
& = L \, \lVert \Pcal_{\j+s\i}(\X-\X') \rVert .
\end{align*}
Moreover, we have from \eqref{equality1} that
\begin{align*}
\sum_{\j \in \J_0} \ \lVert {\Pr}_{\j+s\i}(\X-\X'))  \rVert^2 = \lVert \X-\X' \big\rVert^2.
\end{align*}
Combining the above, we get \eqref{bound}. This establishes part 1 of Lemma \ref{lemma:patchtoimg}. 

Now, let $\D_{\mathrm{patch}}$ be $\theta$-averaged, i.e.,
\begin{equation}
\label{DN}
\D_{\mathrm{patch}} = (1-\theta)\I + \theta \N,
\end{equation}
where $\N$ is nonexpansive. Substituting \eqref{DN} in \eqref{defDi}, and using \eqref{Xidentity}, we can write
\begin{align*}
\D_{\i} =  (1 - \theta) \I + \theta \mathrm{G}_{\i},
\end{align*}
where 
\begin{equation}
\label{Gmn}
\mathrm{G}_{\i} = \sum_{\j \in \J_0} \ \left( \Pcal_{\j+s\i}^* \circ \N \circ \Pcal_{\j+s\i} \right).
\end{equation}
Using \eqref{equality2} and \eqref{equality1} and that $\N$ is nonexpansive, we can show (the calculation is similar to one used earlier) that, for all $\X, \X' \in \R^{q \times q}$,
\begin{align*}
\lVert \mathrm{G}_{\i}(\X) - \mathrm{G}_{\i}(\X') \rVert^2 \leqslant \lVert \X - \X' \rVert^2.
\end{align*}
That is, $\mathrm{G}_{\i}$ is nonexpansive. It follows from \eqref{DN} that $\D_{\i}$ is $\theta$-averaged. This establishes part 2 of Lemma \ref{lemma:patchtoimg}.
\end{proof}

%\end{appendices}

%%===========================================================================================%%
%% If you are submitting to one of the Nature Portfolio journals, using the eJP submission   %%
%% system, please include the references within the manuscript file itself. You may do this  %%
%% by copying the reference list from your .bbl file, paste it into the main manuscript .tex %%
%% file, and delete the associated \verb+\bibliography+ commands.                            %%
%%===========================================================================================%%
\bibliographystyle{sn-aps}
\bibliography{sn-bibliography}% common bib file
%% if required, the content of .bbl file can be included here once bbl is generated
%%\input sn-article.bbl

%% Default %%
%%\input sn-sample-bib.tex%

\end{document}